\documentclass[letterpaper, 10 pt, journal, twoside]{ieeetran}  

\IEEEoverridecommandlockouts                              
                                                          
\usepackage{epsfig}
\usepackage{url}
\usepackage{amsfonts}
\usepackage{amsmath,amssymb,amsthm,bm}

\usepackage{graphicx}
\usepackage{color}
\usepackage{balance} 
\usepackage{algorithm2e}
\usepackage{algorithmic}
\usepackage{graphicx}

\let\labelindent\relax

\usepackage{enumitem}
\usepackage{graphicx} 
\graphicspath{{figure/}}
\usepackage{float}
\usepackage[caption=false,font=footnotesize]{subfig}
\usepackage{soul}
\usepackage{color}

\usepackage{algorithm,algorithmic}
\usepackage{cite} 
\newcommand\highlight[1]{\textcolor{black}{#1}}
\newcommand\highlights[1]{\textcolor{black}{#1}}
\newcommand{\real}[0]{\mathbb R}
\newcommand{\diag}[1]{\ensuremath{\mathrm{diag}(#1)}}

\title{\LARGE \bf \highlight{A Frequency Domain Approach to Predict Power System  Transients }}

\author{Wenqi Cui, Weiwei Yang and Baosen Zhang
\thanks{W. Cui and B. Zhang are with the Department of Electrical and Computer Engineering, University of Washington Seattle, WA 98195, USA	 \{wenqicui, zhangbao\}@uw.edu}%
\thanks{W. Yang is with Microsoft Research, Redmond, WA 98052, USA weiwya@microsoft.com.}%
\thanks{The authors are supported in part by the Climate Change AI.}
}

\begin{document}
\maketitle
\thispagestyle{empty}
\pagestyle{empty}

\begin{abstract}
The dynamics of power grids are governed by a large number of nonlinear differential and algebraic equations (DAEs). To safely operate the system, operators need to check that the states described by these DAEs stay within prescribed limits after various potential faults. However, current numerical solvers of DAEs are often too slow for real-time system operations. In addition, detailed system parameters are often not exactly known. Machine learning approaches have been proposed to reduce the computational efforts, but existing methods generally suffer from overfitting and failures to predict unstable behaviors.  

This paper proposes a novel framework to predict power system \highlight{transients} by learning in the frequency domain. The intuition is that although the system behavior is complex in the time domain, there are relatively few dominate modes in the frequency domain. Therefore, we learn to predict by constructing neural networks with Fourier transform and filtering layers. System topology and fault information are encoded by taking a multi-dimensional Fourier transform, allowing us to leverage the fact that the trajectories are sparse both in time and spatial frequencies. We show that the proposed approach does not need detailed system parameters, greatly speeds up prediction computations and is highly accurate for different fault types. 


\end{abstract}

\section{Introduction}
Increasing the amount of renewable resources integrated in the electric grid is fundamental to reducing carbon emissions and mitigating climate change. Many governments and companies have set ambitious goals to generate their electricity with close to 100\% renewables by 2050~\cite{mai2018electrification}. So far, much of the attention has been paid to increasing the aggregate generation capacities.
However, increased renewable generation capacities also lead to challenging problems
in dynamic stability of the grid~\cite{golden2015curtailment}. A grid can be thought as a large interconnected system of generators, loads and power electronic components, governed by nonlinear differential and algebraic equations (DAEs)~\cite{kundur2017power}. This system also undergos constant disturbances, from load changes to line outages~\cite{sauer2017power}. The predominant goal of power system operations is to make sure that the system stays within acceptable limits under these disturbances. 

\begin{figure}[ht]	
	\centering
	\includegraphics[width=3.2in]{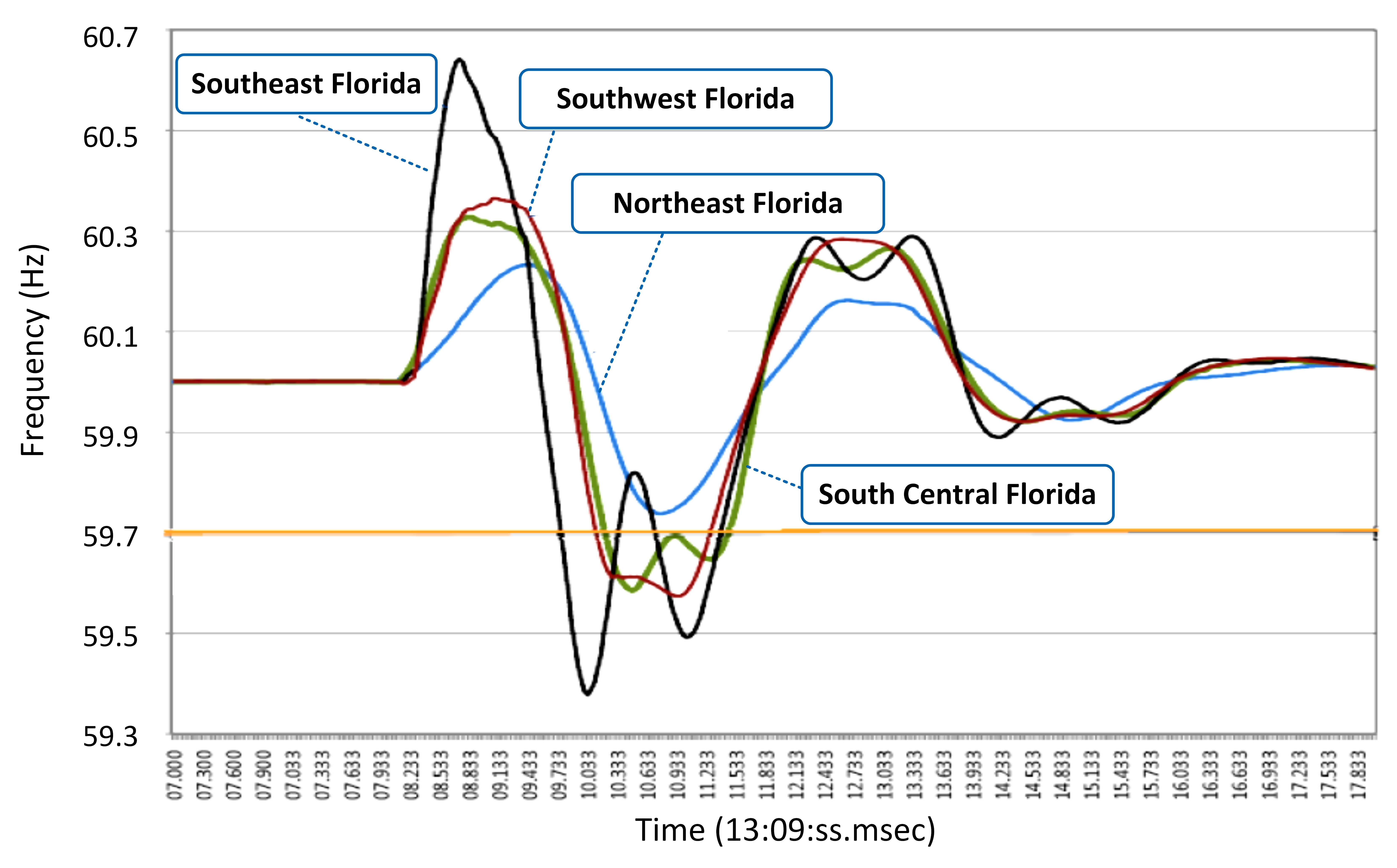}
	\caption{Frequency oscillations during the Florida blackout resulted from a tripped line event in 2008~\cite{florida2008}.}
	\label{fig:florida}
\end{figure}

The system states are governed by different DAEs during the transient across pre-fault, fault-on and post-fault stages~\cite{chiang2011direct}. Fig.~\ref{fig:florida} shows a tripped line in Florida that led to rolling blackouts that impacted the lower two-thirds of state~\cite{florida2008}. Ideally, a system should withstand these types of single events without performance degradation ($N-1$ security)~\cite{sauer2017power}, but this was not the case in Fig.~\ref{fig:florida}. A fundamental reason is that solving the governing DAEs is extremely computationally challenging and not all contingencies can be checked. These DAEs are highly nonlinear and numerical  methods (e.g., implicit and explicit integration) are used to solve them~\cite{kundur1994power}. To study transient stability, numerical algorithms need to use fairly fine discretizations. Even for a moderately sized system, existing numerical solvers may take minutes to simulate only seconds of system trajectories~\cite{huang2017faster}.

\begin{figure}[ht]	
\centering
\includegraphics[width=3.2in]{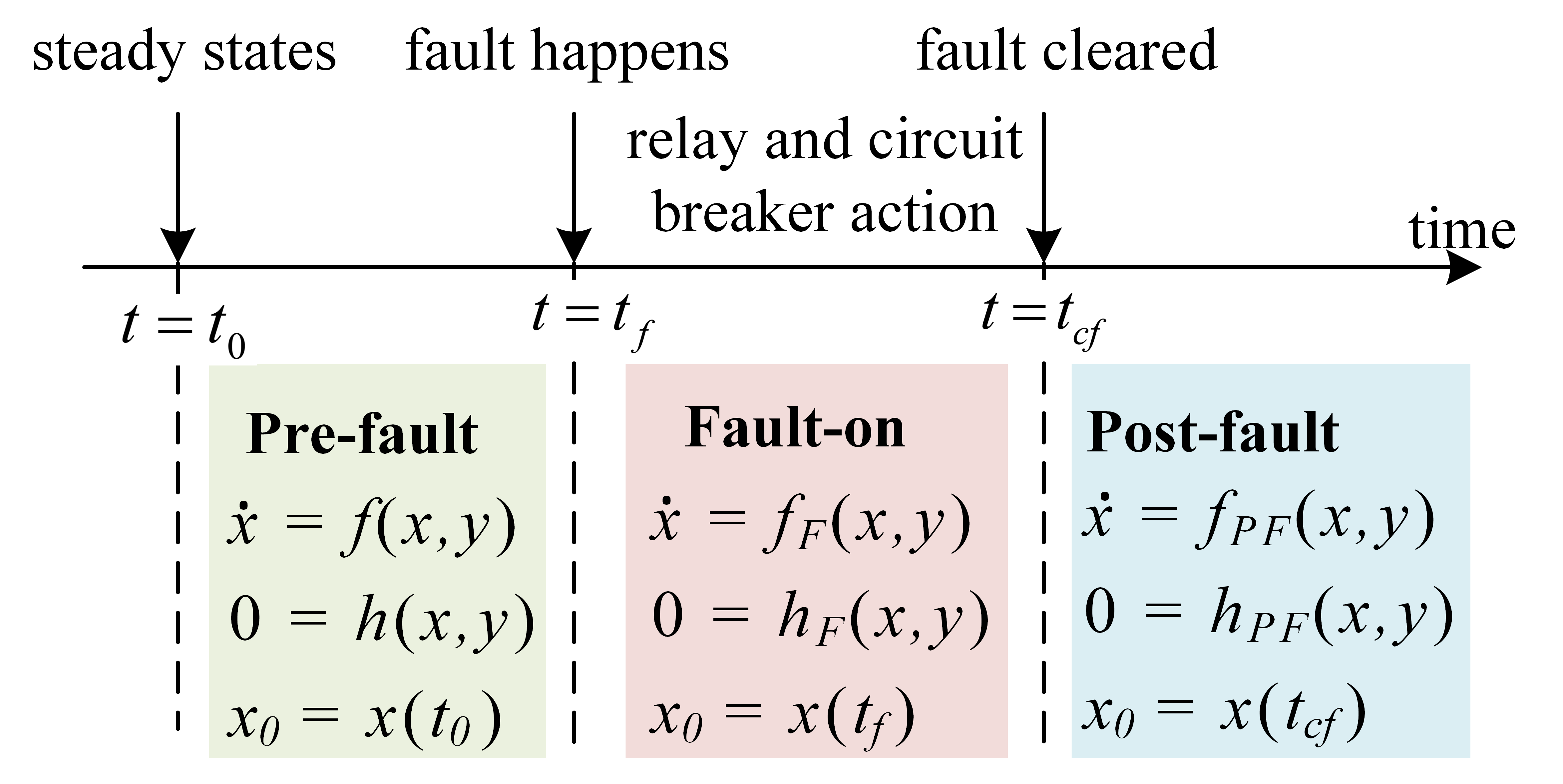}
\caption{Power system operators conduct dynamic power system transient prediction by solving DAEs~\cite{chalishazar2021power, chiang2011direct}, where $x$ and $y$ represent the set of state variables and algebraic variables, respectively. \highlights{The dynamics are described by differential equations $f(\cdot)$ and algebraic equations $h(\cdot)$, with subscript $F$ and $PF$ indicates the fault-on and post-fault period divided by the fault  time $t_f$ and the fault-clear time $t_{cf}$, respectively}. Under certain conditions of power generation, demand and topology, the system trajectory after distrubances is simulated through solving the shifted DAEs.  }
\label{fig:ODE_fault}
\end{figure}

Consequently, only a limited number of scenarios are studied offline and operators tend to restrict the system to operate close to these scenarios. As shown in Fig.~\ref{fig:ODE_fault}, for certain given conditions of power generation, demand and topology, system operators simulate the system trajectories after distrubances through solving the shifted DAEs. By studying scenarios of system conditions offline, system operators will have a checklist about what action needs to be taken in real time to ensure that the systems' states are within the permissible range after critical contingencies. Since renewables have much larger uncertainties than conventional resources, operators often curtail them to artificially limit their generation to avoid operating at ``unknown" regions~\cite{kundur2017power}. For example, some European grids are not allowed to operate at above 40\% wind, no matter how much wind is actually blowing~\cite{wu2019grid}. Therefore, fast and accurate dynamic simulations would greatly increase the actual utilization of renewables in the system. 

Recently, machine learning (ML) approaches have been proposed for dynamical simulation instead of DAE solvers and can reduce the computation time by orders of magnitude. Given the fault information and present measurement of states, ML methods learn feed-forward neural networks to predict the future trajectories. 
\highlight{Most works focus on binary classification to identify whether a system is stable. Long  short-term  memory  based recurrent neural networks are commonly utilized to process sequential data~\cite{ren2022interpretable, zhu2020intelligent}. A shaplet learning approach is proposed to extract spatial-temporal correlations~\cite{ zhu2021networked}. Convolutional neural  networks are adopted in~\cite{yan2019fast, cui2021data} to process data from multiple sources.  }
However, a binary prediction may not provide sufficient information for real-time decision making. For example, to prevent cascading outages, operators need to know the magnitude of the states and the actual trajectories~\cite{chiang2011direct}. And binary predictions are too coarse for making these types of assessments.

To alleviate the limitation of binary prediction, some recent works provide finer trajectory prediction by learning the time domain solutions to the governing DAEs. Polynomial basis are used in~\cite{xia2018galerkin} to approximate the solutions, but the number of basis functions grows exponentially with the system size.  \highlight{Extreme learning machine is utilized in~\cite{su2021intelligent} for online voltage stability margin prediction, but the traces of the system variables are not provided.} Deep neural networks are used in~\cite{yang2019pmu, cepeda2014real} to directly learn to predict the future trajectory from past and current measurements.  However, since power systems are large and sampled sparsely in time, direct regression on time-domain data does not perform very well. A rolling prediction is used in~\cite{yang2019pmu} to improve accuracy, but with a high computational cost. 
\highlight{Instead of the time-domain approach, 
eigenstates of linear swing equations after eigenspace transformation are utilized in~\cite{jalali2022inferring}  to infer system dynamics. However, the assumptions on the linearized system model and uniform damping ratios may not hold for realistic and large-scale systems.}


In addition,  since the majority of trajectories used in training is stable, the learned networks fail predict unstable behaviors. Physics informed neural networks that directly attempt to solve the DAEs have been proposed as an alternative~\cite{stiasny2021learning}, but it does not currently scale beyond small networks.

In this paper, we propose a novel framework for
predicting power system transients
by learning and making predictions in the frequency domain, which provides a computation speed up of more than 400 times compared to existing power system tools. This approach follows the intuition that the system tend to undergo oscillations that have a few dominate temporal-spatial modes. 
We adopt and extend the structure of Fourier Neural Operator in~\cite{li2020fourier} to learn in the frequency domain and recover the time domain trajectories through the inverse Fourier transform. 
Specifically, we design the dataframe to encode the power system topology and fault information, which lead to a 3D Fourier transform. This method is able to make smooth and accurate predictions, capturing both stable and unstable behaviors without the need to manually tune the training data. It improves the MSE prediction error by more than 70\% compared with state-of-the-art AI methods, and vastly improves the detection of unstable behavior.
Code and data are available at https://github.com/Wenqi-Cui/Predict-Power-System-Dynamics-Frequency-Domain.

In summary, the main contributions of the paper are:
\begin{enumerate}
    \item \highlight{We propose a novel machine learning approach to predict transient dynamics in the frequency domain}, which can accurately predict state trajectories based on a few measurements.
    \item We develop a dataframe that encodes spatial-temporal information about the system topology, which greatly reduces the computational complexity in multi-dimensional Fourier transforms.
    \item The time-varying active/reactive power injection and fault-on/clear actions are incorporated in the proposed framework, enabling the prediction of the transients subject to different net power injections and actions.
\end{enumerate}  

The remaining of this paper is organized as follows.  Section~\ref{sec:model} introduces the problem formulation for predicting power system dynamics and transients. Section~\ref{sec:Learning} provides the setup and intuition of learning in the frequency domain. Section~\ref{sec:Frequency} shows the proposed framework  for dynamic transient prediction \highlight{and  Section~\ref{section:FNO} illustrates the construction of dataframe to encode spatial-temporal relationships.}  Section~\ref{sec:simulation} shows the simulation results. Section~\ref{sec:conclusion} concludes the paper.

\section{Model and Problem Formulation} \label{sec:model}

\subsection{Power System Swing Equations}
\begin{figure}[ht]	
\centering
\includegraphics[width=3.5in]{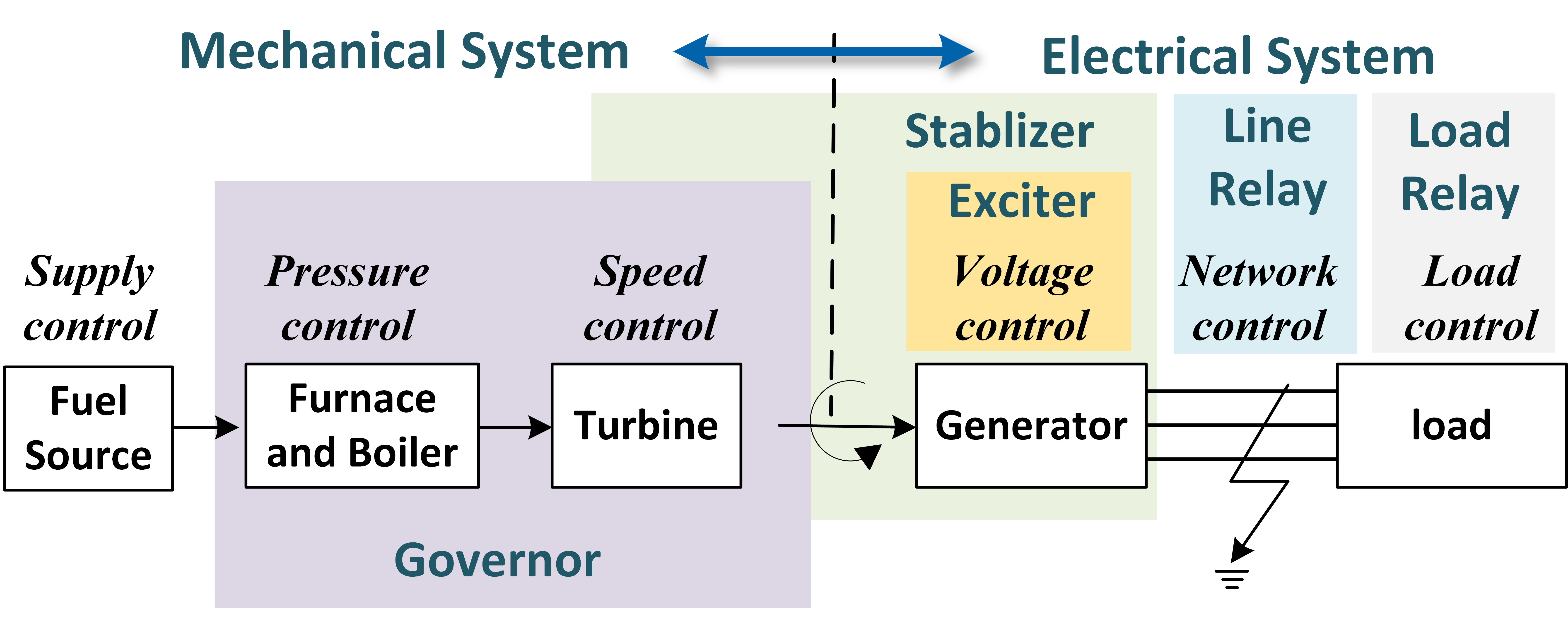}
\caption{Different components contributing to power system transient dynamics~\cite{sauer2017power}. }
\label{fig:Dynamics_Component}
\end{figure}

 The dynamics of power systems depend on the interactions of a myriad of components including governors, exciters, stabilizers, etc., as illustrated in Fig.~\ref{fig:Dynamics_Component}~\cite{sauer2017power}. 
Let $\bm{x} \in \mathbb{R}^{n}$, $\bm{y} \in \mathbb{R}^{m}$,  $\bm{a} \in \mathbb{R}^{d}$ be all the state variables, algebraic variables and external input variables, respectively.\footnote{Throughout this paper, vectors are denoted in lower case bold and matrices are denoted in upper case bold, while scalars are unbolded. If not specified, all the vectors are time-varying. The vectors followed by $(t)$ denotes the value at the time $t$. } The complete power system model for calculating system dynamic response relative to a disturbance can be described by a set of DAEs as follows~\cite{xia2018galerkin, chiang2011direct}:
\begin{equation}\label{eq:Dynamic}
\left\{\begin{array}{c}
\dot{\bm{x}}=\bm{f}(\bm{x}, \bm{y}, \bm{a}) \\
\bm{0}=\bm{h}(\bm{x}, \bm{y}, \bm{a})
\end{array}\right.
\end{equation}
where the differential
equation  $\bm{f}: \mathbb{R}^{n} \times \mathbb{R}^{m} \times \mathbb{R}^{d} \rightarrow \mathbb{R}^{n}$ typically describes the internal dynamics of devices such as the speed and angle of generator rotors, the response of generator control systems (e.g., excitation systems, turbines, governors), the
dynamics of equipment including DC lines, dynamically modeled loads 
and their control systems. Correspondingly,  $\boldsymbol{x} \in \mathbb{R}^{n}$ is the state variables such as generator rotor angles, generator velocity deviations (speeds), electromagnetic flux, various control system internal variables, etc. The set of algebraic equations $\bm{h}: \mathbb{R}^{n} \times \mathbb{R}^{m} \times \mathbb{R}^{d} \rightarrow \mathbb{R}^{m}$  describes the electrical transmission system and interface equations.  Correspondingly, 
$\bm{y} \in\mathbb{R}^{m}$ is the algebraic variable such as voltage magnitude and angles.

The external input variables $\bm{a}\in \mathbb{R}^{d}$ acting on the equations are power injection from generators, automatic generation control
systems, fault-response actions and so on~\cite{chiang2011direct, jia2021transient}. In this paper, we mainly consider the  uncertainties in power generation and demand, as well as the fault-response actions $\bm{u}$. Let the active and reactive net power injection be $\bm{p}$ and $ \bm{q}$, then the
external input variables  is sometimes written as the tuple $\bm{a}=(\bm{p}, \bm{q}, \bm{u})$.

Under changing generator and load conditions, power systems are operated to withstand the occurrence of certain contingencies. To ensure that cascading outages will not occur for the set of critical disturbances, the state variables need to stay within permissible ranges during the transient process of power system after disturbances~\cite{sauer2021power, chiang2011direct}. 
  For even a moderate power system with tens or hundreds of buses, it may be governed by hundreds or thousands of DAEs.


\subsection{Transient Dynamics}
Disturbances lead to deviations in the states $\bm{x}$ and variables $\bm{y}$ through a step change of parameters in~\eqref{eq:Dynamic}. For example, a  short  circuit  on  a  transmission line
will results in a sudden change of the susceptance and conductance in set of algebraic equations,
 depending on the specific fault types (e.g., single-phase-to-ground, two-phase-to-ground, line-to-line, three-phase-to-ground, etc.).  

Suppose a fault happens at the time $t_f$ and is cleared at the time $t_{cl}$.~\footnote{The  disturbance such as a short circuit on a transmission line is automatically cleared by protective relay operation after a certain amount of time.}  The pre-fault stage is defined as the period before the fault happens at $t_f$. The system evolves from the initial state $\bm{x}(t_0)$ as:

\begin{equation}\label{eq:pre_fault}
\left\{\begin{array}{l}
\dot{\bm{x}}=\bm{f}(\bm{x}, \bm{y},\bm{a}; \bm{x}(t_0)) \\
\bm{0}=\bm{h}(\bm{x}, \bm{y}, \bm{a})
\end{array}\right., \quad t_0\leq t<t_f
\end{equation}

The sudden parameter changes after disturbances will lead to a shift of the swing equations. 
The fault-on system evolves with $k$ subsequent actions from system relays and circuit breaks. Suppose the $j$-th action is taken at $t_{F,j}$, the fault-on system is described by several set of equations~\cite{chiang2011direct}


\begin{equation}\label{eq:on_fault}
\begin{array}{ll}
    \left\{\begin{array}{l}
    \dot{\bm{x}}=\bm{f}_{F,1}(\bm{x}, \bm{y},\bm{a};\bm{x}(t_f)) \\
    \bm{0}=\bm{h}_{F,1}(\bm{x}, \bm{y},\bm{a} )
    \end{array}\right.
    , & t_f\leq t<t_{F,1}\\
    \quad\quad\quad\vdots \\
    \left\{\begin{array}{l}
    \dot{\bm{x}}=\bm{f}_{F,k}(\bm{x}, \bm{y},\bm{a};\bm{x}(t_{F,k})) \\
    \bm{0}=\bm{h}_{F,k}(\bm{x}, \bm{y},\bm{a} )
    \end{array}\right.
    , & t_{F,k-1}\leq t<t_{cl}
\end{array}
\end{equation}


The post-fault stage refers to the system after the fault is cleared. 
\highlights{The post-fault system evolves with the differential equation  starting from the post-fault initial state $\bm{x}(t_{cl})$, written as~\cite{chiang2011direct}}

\begin{equation}\label{eq:post_fault}
\left\{\begin{array}{l}
\dot{\bm{x}}=\bm{f}_{PF}(\bm{x}, \bm{y},\bm{a}; \bm{x}(t_{cl})) \\
\bm{0}=\bm{h}_{PF}(\bm{x}, \bm{y}, \bm{a})
\end{array}\right., \quad t\geq t_{cl}.
\end{equation}


Despite of large numbers of variables $\bm{x}$ and $\bm{y}$, not all of them are observable to system operators.  For a system with $N$ buses, typically the main variables of interest are  the angle $\delta_i$, rotor angle speeds (frequency) deviation $\omega_i$ and voltage $v_i$ at each bus $i$.  
Denote $\left[N\right]:=\{1,\dots, N\}$, $\bm{\delta}:=\left(\delta_i, i \in \left[N\right] \right) \in \mathbb{R}^N$, $\bm{\omega}:=\left(\omega_i, i \in \left[N\right] \right) \in \mathbb{R}^N$, $\bm{v}:=\left(v_i, i \in \left[N\right] \right) \in \mathbb{R}^N$.
These variables of interest are described by a three-tuple, denoted by $\bm{s}=(\bm{\delta}, \bm{\omega}, \bm{v})\in \mathbb{R}^{3N}$. Note that other variables can also be included in $\bm{s}$ if they are observable. The key to safe dynamic operation of power system is to predict the future of the system trajectories, given the fault information, some observations $\bm{s}$ and the expected clearing actions  $\bm{u}$. Based on these trajectories, interim actions  like load shedding or emergency generation can be taken to reduce the impact of the faults~\cite{kundur1994power, sauer2017power}.

\subsection{Current Approaches and Challenges}
Current approaches in power system dynamic prediction is based on solving~\eqref{eq:pre_fault}-\eqref{eq:post_fault}, which are highly nonlinear equations.
System operators typically rely on numerical integration, such as Runge-Kutta (RK) methods or trapezoidal rule, to iteratively approximate the solution of~\eqref{eq:pre_fault}-\eqref{eq:post_fault} in small time intervals~\cite{kundur1994power}. 
However, because of the highly nonlinear nature of the DAEs, very fine discretization steps are required for these numerical methods. As a result, these approaches may be too slow for real-time decision-making. \highlight{Some solvers use reduced-order model and convert DAEs to ordinary differential equations (ODEs) to simulate the dynamic response of generators~\cite{chow1992toolbox}. } For a moderately sized system, existing numerical solvers take minutes to simulate only seconds of system trajectories~\cite{huang2017faster}.
As an alternative, system operators also use manual heuristics to  take  actions,  but  this  strategy  is becoming less tenable as  renewables introduce distinctly different operating scenarios. 


    
    

\section{Learning for Dynamic Transient Predictions}\label{sec:Learning}


\subsection{Problem Setup}
Learning-based approaches try to find a mapping from present measurements to future trajectories. The predictions are then obtained through function evaluations, which significantly reduces the computational time compared to the conventional numerical approaches.

\begin{figure}[H]
    \centering
    \includegraphics[width=3in]{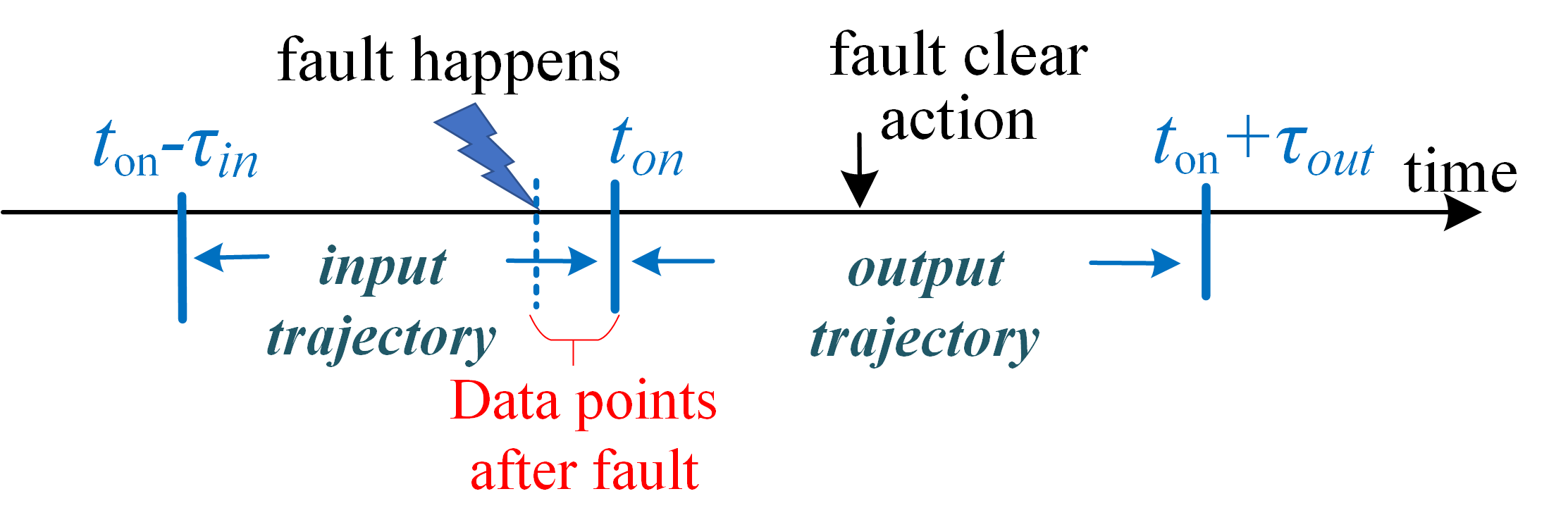}
    \caption{\highlight{Illustration of the trajectory prediction starting at the time stamp $t_{on}$ for $\tau_{out}$ number of time steps, using  $\tau_{in}$ observations.}
 }
    \label{fig:prediction}
\end{figure}

The problem we are interested in is to predict the trajectory of the states $\bm{s}$ starting at the time stamp $t_{on}$ for $\tau_{out}$ number of time steps 
\highlight{with the sampling interval $\Delta t$, as illustrated in Fig. \ref{fig:prediction}}. The input are $\tau_{in}$ observations of the states from  $\bm{s}(t_{on}-\tau_{in})$ to $\bm{s}(t_{on}-1)$, the external inputs   from $\bm{a}(t_{on}-\tau_{in})$ to $\bm{a}(t_{on}-1)$, and the expected fault-clear actions from $\bm{u}(t_{on})$ to $\bm{u}(t_{on}+\tau_{out}-1)$. We write $\bm{s}_{in}=(\bm{s}(t_{on}-\tau_{in}),\cdots,\bm{s}(t_{on}-1))$, $\bm{a}=(\bm{a}(t_{on}-\tau_{in}),\cdots,\bm{a}(t_{on}-1))$, $\bm{u}_{out}=(\bm{u}(t_{on}),\cdots,\bm{u}(t_{on}+\tau_{out}-1))$   and $\bm{s}_{out}=(\bm{s}(t_{on}),\cdots,\bm{s}(t_{on}+\tau_{out}-1))$.

Our goal is to find  a mapping $G$ from the space of input ($\bm{s}_{in}$, $\bm{a}$, $\bm{u}_{out}$) to output trajectories $\bm{s}_{out}$. In this paper we consider the mapping realized through (deep) neural networks with parameters $\bm{\Phi}$. The prediction is then given by
\begin{equation}\label{eq:mapping}
    \hat{\bm{s}}_{out} = G_{\bm{\Phi}}(\bm{s}_{in},\bm{a},\bm{u}_{out}).
\end{equation}
The exact form of $\bm{\Phi}$ depends on the  structure of neural network used. In this paper, we adopt the  Fourier neural operator to learn in the frequency domain, and details will be specified later in Section~\ref{sec:Frequency}.

Let $H$ be the batch size and $\hat{\bm{s}}_{out}^{i}$ be the prediction of the $i$-th sample for $i=1,\cdots,H$. The weights of neural network $\bm{\Phi}$ are updated by back-propagation to minimize loss function $L(\bm{\Phi})$ defined by the mean absolute percentage error
(MAPE) between predicted trajectory and the actual trajectory $    L(\bm{\Phi})=\frac{1}{H}\sum_{i=1}^{H}\frac{||\hat{\bm{s}}_{out}^i-\bm{s}_{out}^{i}||_1}{||\bm{s}_{out}^{i}||_1}$~\cite{li2020fourier}. 


\subsection{Current ML Approaches and Limitations}
\begin{figure}[ht]	
	\centering
	\includegraphics[width=3.2in]{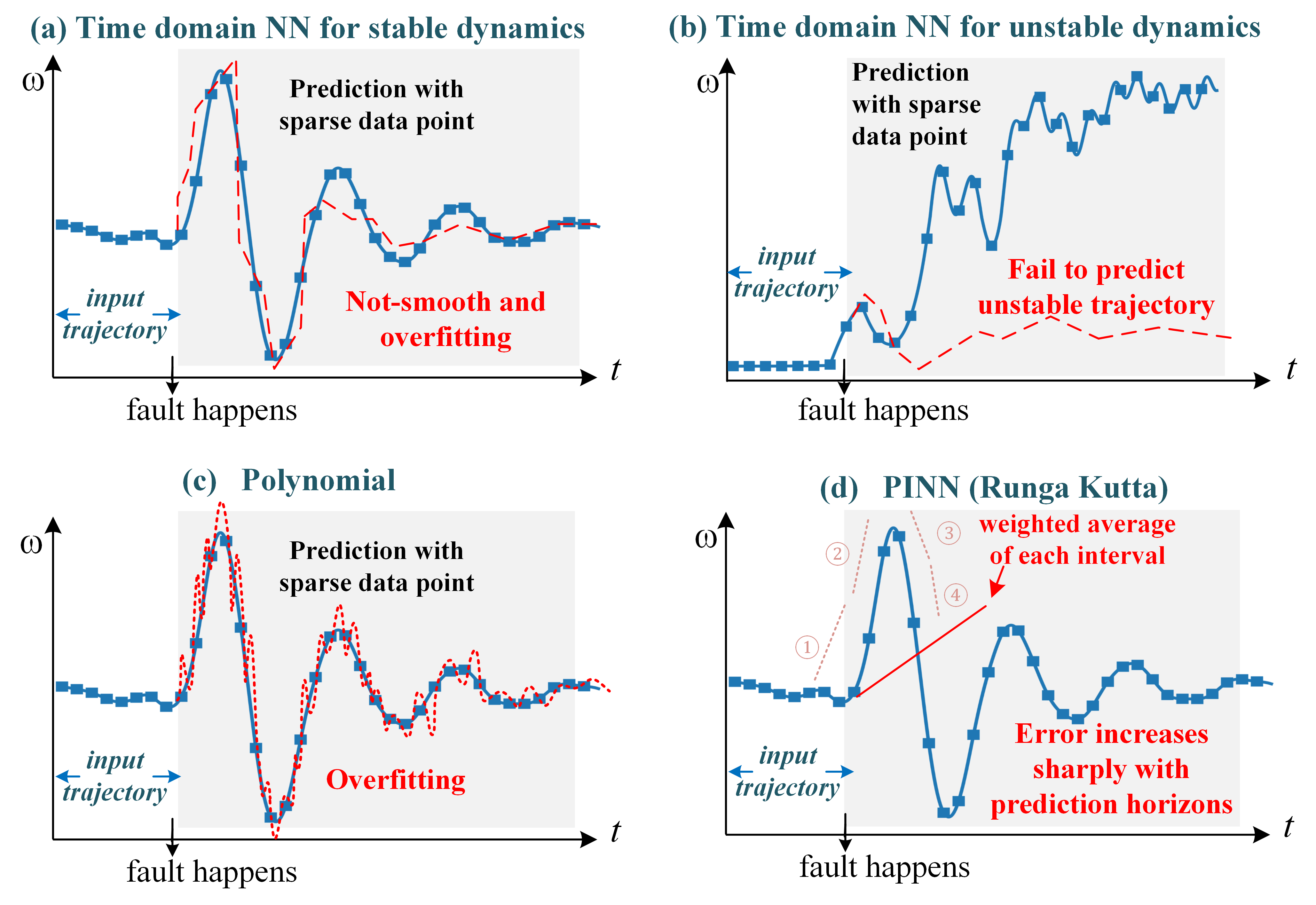}
	\caption{Illustration of power system transient prediction with existing machine learning approaches. (a) Purely learning in the time domain tend to overfit. (b)  Since the vast majority of historical data are stable, a ML method tends to not predict unstable trajectories. (c) Fitting the nonlinear dynamics with polynomial basis also easily lead to over-fitting. (d) The accuracy of PINN decreases sharply and fails to provide meaningful results for longer horizons.  }
	\label{fig:Intuition_ML}
    \end{figure}

Learning the power system transient dynamics is not trivial because states undergo nonlinear oscillations. Predictions with three existing approaches  are illustrated in Fig.~\ref{fig:Intuition_ML}. The blue line is the trajectory of the frequency deviation on a bus before and after a fault. The blue squares are true trajectory sampled at discrete times  and the grey area is the prediction horizon.

A standard approach is to use a neural network to learn the time-domain mapping from the input to output. 
As illustrated in Fig.~\ref{fig:Intuition_ML}(a) and Fig.~\ref{fig:Intuition_ML}(b),  purely learning in the time domain will easily overfit and cannot learn a smooth curve like the true trajectories. More importantly, generic machine learning approaches prone to false negative errors. Since the vast majority of historical trajectories are stable, a ML method tends to not predict unstable trajectories. 
This would lead to catastrophic consequences if the system operator does not take actions to mitigate instabilities.

Similarly, fitting the nonlinear dynamics with polynomial basis will also easily lead to over-fitting, as  illustrated in Fig.~\ref{fig:Intuition_ML}(c).  Recently, Physics-Informed Neural Networks have been proposed in~\cite{raissi2019physics} to learn solutions that satisfy equations from implicit Runge-Kutta (RK) integration. This approach has been applied to  power system swing dynamics in~\cite{stiasny2021learning}. Since RK method is the weighted sum of ODE solutions in discretized intervals, its accuracy decreases sharply when predicting trajectories with large oscillations for a longer horizon (e.g., larger than 1 second), as illustrated in Fig.~\ref{fig:Intuition_ML}(d) (the prediction errors are larger than then limit of the y-axis).

\subsection{The Proposed Approach: System Dynamics in the Frequency Domain}

Because of the above challenges when learning in the time domain, we propose a new approach for learning power system  transient dynamics in the frequency domain. Here we use a simple swing equation model for transient dynamics to illustrate the intuition for this approach~\cite{kundur1994power, guo2018graph}:
\begin{subequations}\label{eq:Linear_Dynamic}
	\begin{align}
	&\dot{\delta_i} =\omega_i\\
	\begin{split}
	&M_i\dot{\omega}_{i} =p_{i}-D_i \omega_i -\sum_{j=1, j\neq i}^{N}B_{i j}(\delta_i-\delta_j)
	\end{split}
	\end{align}
\end{subequations}
where $ i\in\left[N\right]:=\{1,\dots, N\} $ is the index of buses, $\boldsymbol{M} := \diag{M_i, i \in \left[N\right]} \in \real^{N \times N}$ are the generator inertia constants, $\boldsymbol{D} := \diag{D_i, i \in \left[N\right]} \in \real^{N \times N}$ are damping coefficients, $\boldsymbol{p}:=\left(p_{i}, i \in \left[N\right] \right) \in \real^N$ are the net power injections, $\boldsymbol{B}:=\left[B_{ij}\right]\in \real^{N \times N}$ is the susceptance matrix. 

With the assumption that  the inertia and
damping of the buses are proportional to its power ratings (i.e,  $D_i/M_i=\gamma$ for all $i\in[N]$), \eqref{eq:Linear_Dynamic}  can be explictly solved~\cite{guo2018graph}. 
Let  $\bm{C}$ be the  incidence matrix. The (scaled) graph Laplacian matrix  is  $\bm{\Gamma }=\bm{M}^{-1 / 2} \bm{C} \bm{B} \bm{C}^\top \bm{M}^{-1 / 2}$, where $0=\lambda_{1}<\lambda_{2} \leq \cdots \leq \lambda_{n}$ are the eigenvalues with corresponding orthonormal eigenvectors $
\bm{r}_{1}, \bm{r}_{2}, \ldots, \bm{r}_{n}$. 
Suppose there is a step change $\Delta\bm{ p}$  in the net power injection, and its decomposition along the eigenvectors is $\Delta\bm{ p}=\sum_{i} \hat{p}_{i} \bm{M}^{1 / 2} \bm{r}_{i}$. Then, the solution of equations~\eqref{eq:Linear_Dynamic} is~\cite{guo2018graph}
\begin{equation}
\bm{\omega}(t)=\sum_{i=1}^{n} \frac{\hat{p}_{i}}{\sqrt{\gamma^{2}-4 \lambda_{i}}}\left(e^{\phi_{i,+} t}-e^{\phi_{i,-} t}\right) \bm{M}^{-1 / 2} \bm{r}_{i}
\end{equation}
where $
\phi_{i,+}:=\frac{-\gamma+\sqrt{\gamma^{2}-4 \lambda_{i}}}{2}
\quad \phi_{i,-}:=\frac{-\gamma-\sqrt{\gamma^{2}-4 \lambda_{i}}}{2}.$

Note that $\phi_{i,+}$ are complex numbers with non-zero imaginary part if
 $\gamma^{2}-4 \lambda_{i}<0$, which results in  sinusoidal oscillations. Thus,  the sinusoidal basis in Fourier transform (and inverse Fourier transform) is a natural fit for  power system transient dynamics. 
Because of  the finite set of eigenvalues, there is a finite number of modes in the frequency domain. As a result, the trajectories are sparse in the frequency domain, making it easier to learn after Fourier transform.

\highlight{However, the analysis based on the linear model~\eqref{eq:Linear_Dynamic}
cannot be applied to more realistic systems as illustrated in Fig.~\ref{fig:Dynamics_Component}, where high-order nonlinear differential equations are involved. This is the reason why we need learning to predict the transient dynamics. In the next sections, we will show the framework of learning in the frequency domain and conduct numerical verification using full-order  model for system dynamics.}



\section{Learning in the Frequency Domain}\label{sec:Frequency}

\subsection{Structure of the Neural Network}
\begin{figure}[H]
    \centering
    \includegraphics[width=3.5in]{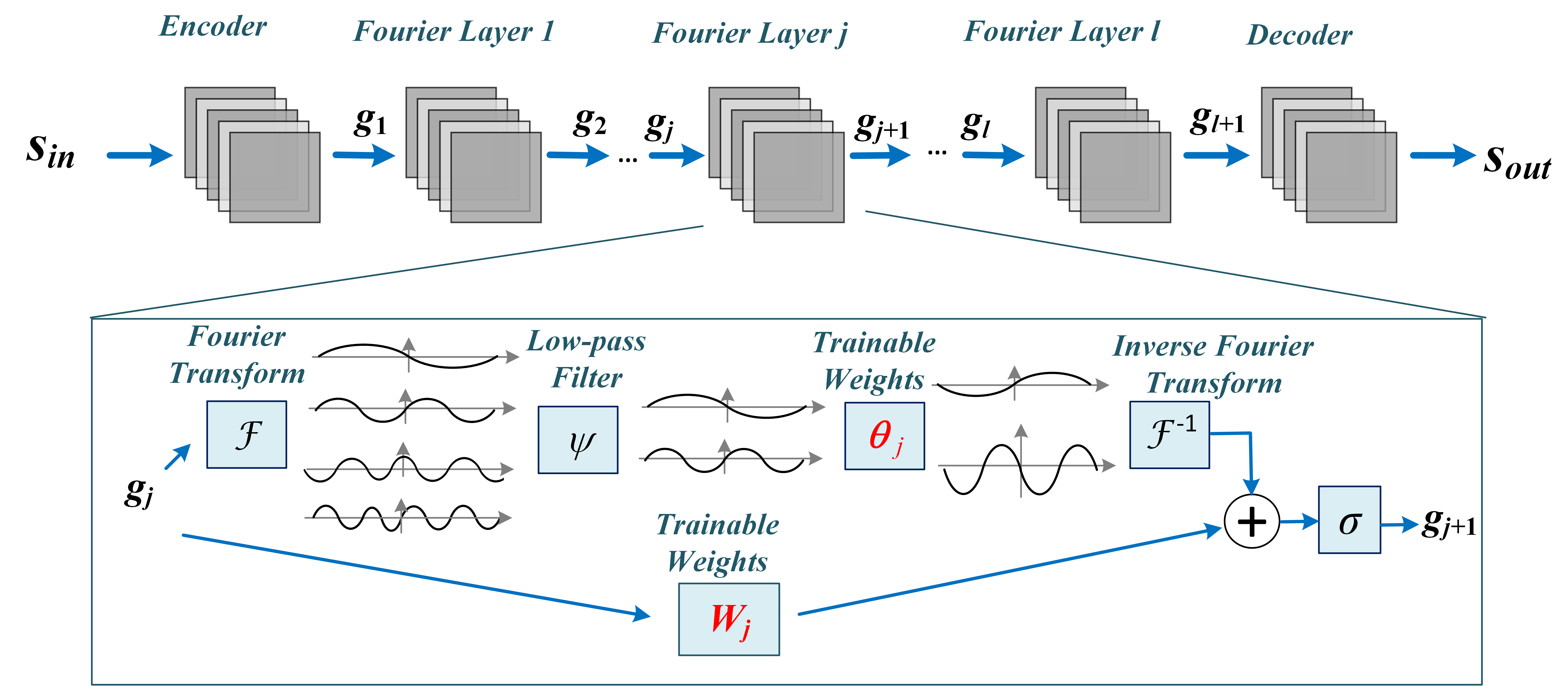}
    \caption{\highlight{The structure of neural networks for power system transient prediction using frequency-domain information. The input trajectory is first encoded using the framework introduced later in Section V. The encoded data is then passed through several Fourier Layers~\cite{li2020fourier}, where the input and output of the $j$-th layer is $\bm{g}_{j}$ and $\bm{g}_{j+1}$, respectively.
    The $j$-th layer consists of trainable weights $\bm{\theta}_{j}$ for learning in the Fourier domain, and trainable weights $\bm{W}_{j}$ to keep the track of aperiodic and distorted
waveform.}
 }
    \label{fig:FNO_structure_total}
\end{figure}

\highlight{We construct the structure of neural network shown in Fig.~\ref{fig:FNO_structure_total}, which consists of several Fourier layers for learning in the frequency domain. The input trajectory is first passed through an encoder to integrate the spatial-temporal relationships and the fault information. 
 }
 The encoded data is then passed through $l$ level of Fourier Layers~\cite{li2020fourier}, where the input of the $j$-th layer is $\bm{g}_{j}$ and the output is $\bm{g}_{j+1}$ for $j=1,\cdots,l$. \highlight{Each Fourier layer consists of one path with  trainable weights $\bm{\theta}_{j}$ that  learns periodic components in the frequency domain, and another path with trainable weights $\bm{W}_{j}$  that directly operate on the time-domain data.  Intuitively, the second path (often called the pass-through layer in Machine Learning literature) with weights $\bm{W}_{j}$ helps to keep the track of aperiodic and high-frequency component.}
 \highlight{In the following, we illustrate the structure of each Fourier Layer.  The dimension for each tensor will be specified later after we elaborate the encoder  in Section~\ref{section:FNO}.
}


\subsection{Fourier Layer}
For the input of each layer, we conduct discrete Fourier transform $\mathcal{F}$ to convert the input trajectory into the frequency domain~\cite{brunton2019data}. Inspired by the work in~\cite{li2020fourier}, we use neural networks parameterized by $\bm{\theta}_j$ to learn in the frequency  in each layer $j$, and then recover the time-domain sequences by inverse Fourier transform ${\mathcal{F}}^{-1}$. This process is defined as Fourier neural operator $\mathcal{K}_{\bm{\theta}_j}(\cdot)$ represented by 
\begin{equation}\label{eq: Fourier_operator}
\mathcal{K}_{\bm{\theta}_j}\left( \bm{g}_{j}\right)={\mathcal{F}}^{-1}\left(\bm{\theta}_j \cdot\psi \left(\mathcal{F} \bm{g}_{j}\right)\right),
\end{equation}
where the function $\psi(\cdot)$ is a low-pass filter that truncates the Fourier series at a maximum number of modes $k_{\max }$
for efficient computation~\cite{li2020fourier}. Then $\bm{\theta}_j$ is the weight tensor \highlight{that} conducts linear combination of the modes in the frequency domain.


The output of the $j$-th layer adds up Fourier neural operator with the initial time-domain sequence weighted by $\bm{W}_{j}$ to recover aperiodic and high-frequency components
\begin{equation}\label{eq: Fourier_layer}
\bm{g}_{j+1}=\sigma\left(\bm{W}_{j} \bm{g}_{j}+\mathcal{K}_{\bm{\theta}_j}\left( \bm{g}_{j}\right)\right),
\end{equation}
where $\sigma$ is a nonlinear activation function whose action is defined component-wise. We use ReLU in this paper.

\highlight{ Even though we limit  at most $k_{\max}$ Fourier modes after the low-pass filter $\psi$ after the Fourier transform,  the linear transform $\bm{W}_{j}$ maintains high-frequency modes.}
\highlight{The cut-off frequency $k_{\max}$  of the low-pass filter is a tradeoff between the number of frequency component that are kept versus the computational complexity. If too few frequencies are selected, there is not enough information in the frequency domain to learn well. If too many frequencies are selected, then we need to learn a high dimensional set of weights, negating the benefit of learning in the frequency domain. The trade-off we adopted is to keep a small number of modes in the frequency domain and pass them through nonlinear layers, while still using a direct path as shown in Fig.~\ref{fig:FNO_structure_total}. Intuitively, this means that the low frequency modes should be learned in frequency domain, while the higher frequency modes can be directly handled using the time domain signal. The exact value of $k_{\max}$ involves some trial and error. We will show the numerical study about the effect of $k_{\max}$ in Section~\ref{sec:simulation}.}

\subsection{Multi-Dimensional Fourier Transforms}

The above approach of learning the weights in the frequency domain and recovering the trajectory with inverse Fourier transform provides the advantage in fitting oscillatory functions, by learning smooth curvatures and avoiding over-fitting. For a system with $N$ buses, there are $3N$ state variables we are interested: the voltage, angle and frequency at each bus.    
However, conducting Fourier transform with $3N$ dimensions is time consuming even for moderatedly sized power systems.

Another choice is to neglect the dependence and purely conduct 1D Fourier transform on the time dimension.  However, this will degrade  the prediction performance since the networked structure is an important cause of the oscillations. Moreover, the time varying parameters and the fault-clear actions should be considered as well.

 To overcome these challenges, we design a novel dataframe that encodes time-varying parameters, fault information and spatial-temporal relationships in transient dynamics in the next section.

\section{Encoding Spatial-Temporal Relationships} \label{section:FNO}


\begin{figure*}[h]
    \centering
    \includegraphics[width=\textwidth]{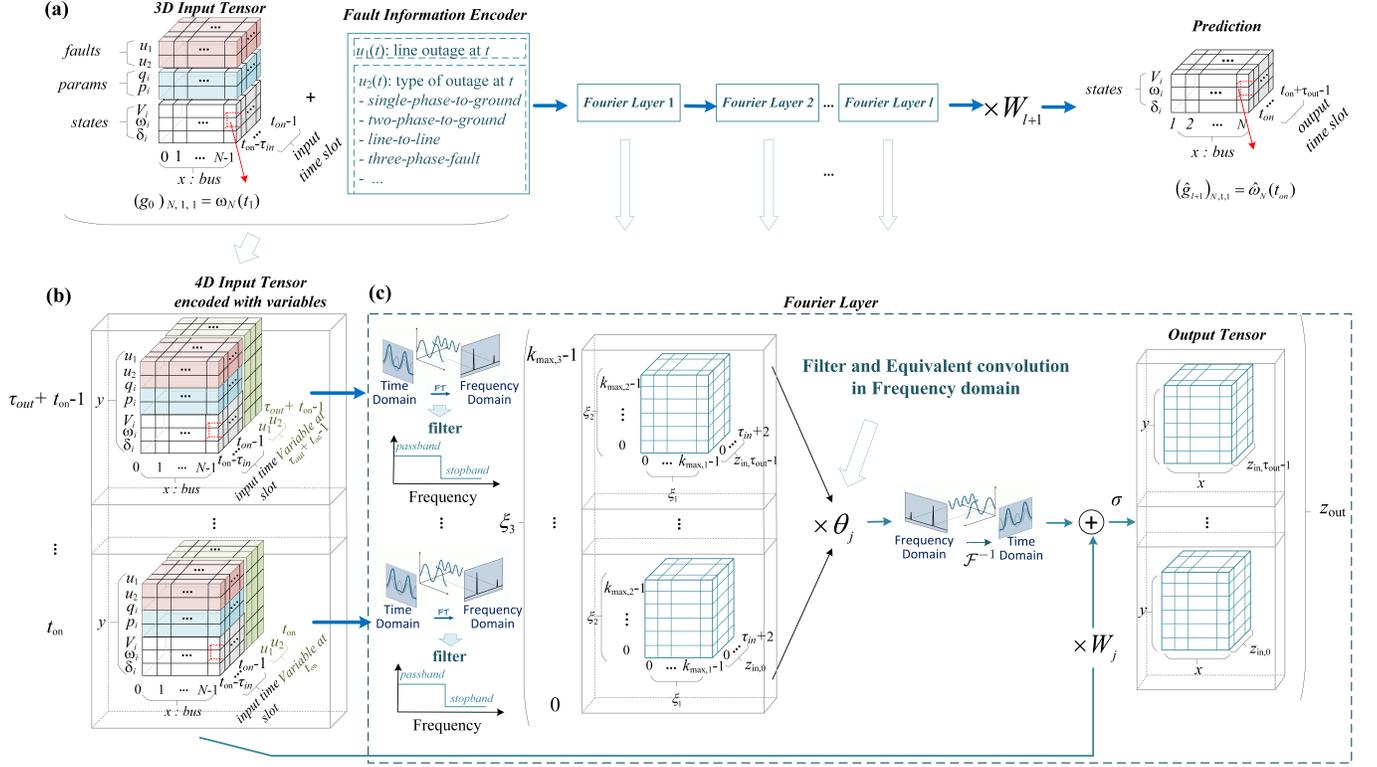}
    \caption{Structure of power system transient prediction using Fourier Neural Operator (a) The data frame that incorporate time-varying parameters and fault-on/clear actions within input time slots. (b) Incorporation of future fault-clear actions in the output time slots. (c) Fourier layer for Learning in the the frequency domain~\cite{li2020fourier}. }
    \label{fig:FNO_structure}
\end{figure*}

\subsection{Spatial-Temporal Relationship in Transient Dynamics}
We construct 3D tensors to encode the input trajectories such that the spatial-temporal relationships in the power system can be included. In addition, computation complexities of  Fourier transforms are also reduced.  The proposed framework is shown in Fig.~\ref{fig:FNO_structure}(a). \highlight{Each data point is indexed by three dimensions and written as  $\left(\bm{g}_0\right)_{x,y,z_{in}}$.}
The $x$ axis indexes the buses from $0$ to $N-1$. The $y$ axis indexes different state variables for each bus with $y = 0,1,2$ stands for $\delta_x$, $\omega_x$,  $V_x$, respectively. The $z_{in}$ axis is for input time interval and $z_{in}=t_{on}-\tau_{in},\cdots,t_{on}-1$.
\highlight{For example, $\left(\bm{g}_0\right)_{5,1,10}$ is the frequency  at bus $5$ at the time step $10$, and $\left(\bm{g}_0\right)_{6,2,10}$ is the voltage at bus $6$ at the time step $10$. }

\highlight{ We would like to highlight that the spatial information in this manuscript can be understood as the inherent spatial similarity of signals. For example, it is well known that the angle speed of generators tend to appear as groups that swing in similar patterns~\cite{zhao2014design}.
This indicates the correlation along the axis of bus indexes. Moreover, the angle, angle speed and voltage also typically oscillate at similar frequency. This indicates  the correlation along the axis of the type of signals. The Fourier transform is most commonly defined  along the axis of time, namely, the 1D Fourier transform along the time series. In this paper, we conduct 3D Fourier transform along the dimension of $x,y$ and the time horizon. This way, the correlations along different buses, along type of signals and along the time horizon are extracted. }

\highlight{\subsection{ Encoding On-Fault and Fault-Clear Information}}

Importantly, we aim to predict the trajectories under changing net injections and faults. This is different from most previous works that learn a static mapping from input to output time sequences for fixed parameters. Therefore, we encode parameters and fault-clear actions explicitly in the input tensor as shown in Fig.~\ref{fig:FNO_structure}(a). \highlight{Note that the fault-clear action may not known in advance, and we set up an expected relay time to predict the dynamic behaviors.
The aim is to  use the learning-based method to reproduce the results from solvers. That is, given the relay actions after the fault happens, the solvers can compute the trajectories after the fault. Likewise, we envision the relay actions as an extra input encoded in the data frame.}

\highlight{The fault information is encoded in $u_1(t)$ and $u_2(t)$ , which are variables that contain the location of fault and the type of fault at the time step $t$, respectively. For example, suppose the fault at the time step $z_{out}+t_{on}$ includes the trip of line 100 and the fault type is line-to-line fault. We encode the fault location as  $100$ and this fault type as  $20$. Then,  $u_1(z_{out}+t_{on})=100$ and $u_2(z_{out}+t_{on})=20$, respectively. 
For the time step $z_{out}+t_{on}$ between the line tripping and line relay,  $u_1(z_{out}+t_{on})\neq 0$ and $u_2(z_{out}+t_{on})\neq 0$. If there is no fault happening  at the time step $z_{out}+t_{on}$, then $u_2(z_{out}+t_{on})=0$ and $u_2(z_{out}+t_{on})=0$.  Hence, the information of line tripping and relay is  inherently included when we add $u_1(t)$ and $u_2(t)$ to the data frame for $t=t_{on}-\tau_{in},\cdots, t_{on}+\tau_{out}-1$. Next, we show how to attach $\bm{u}$ and net injections to the data-frame.}

\subsection{Encoding Time-Varying Parameters and Actions}
Time-varying parameters   include the net active power injection $\bm{p}(t)=(p_1(t),\cdots,p_N(t))$ and reactive power $\bm{q}(t)=(q_1(t),\cdots,q_N(t))$. We stack them on the y axis on $y=3$ and $y=4$ as shown in blue part of Fig.~\ref{fig:FNO_structure}(a). The benefit of this design is that the variance of $\bm{p}(t)$ and $\bm{q}(t)$ through time is naturally incorporated in the $z_{in}$ axis.
The fault information $u_1(t)$ and $u_2(t)$ for $t=t_{on}-\tau_{in},\cdots,t_{on}-1$ are stacked to the y-axis as $y=5$ and $y=6$, shown in red part of Fig.~\ref{fig:FNO_structure}(a).

Fault-clear actions may happen in the predicted time horizon $\left [ t_{on}, t_{on}+\tau_{out}-1\right ]$. To incorporate future actions and temporal dependence in the prediction time steps, we expand the 3D input tensor in Fig.~\ref{fig:FNO_structure}(b) along the output time sequence, with the new axis $z_{out}=0,\cdots,\tau_{out}-1$ correspond to the time stamp from $t_{on}$ to $t_{on}+\tau_{out}-1$. \highlight{This is visualized in the green part of the Tensor in Fig.~\ref{fig:FNO_structure}(b), where the axis of $z_{in}$ is attached with $z_{out}+t_{on}$, $u_1(z_{out}+t_{on})$ and $u_2(z_{out}+t_{on})$, respectively. Each data point in the 4D tensor is written as $\left(\bm{g}_1\right)_{z_{out}, x,y,z_{in}}$, where $x$ indexes bus, $y$ indexes the type of signals (e.g., $y=0,1,2$ corresponds to the angle, angle speed, and voltage, respectively), and  $z_{out}$ indexes the output time steps. The index $z_{in}=0,\cdots,t_{on}-1$ corresponds to the time stamp of the input trajectory. The index $z_{in}=t_{on}$ stands for the location of the fault,  $z_{in}=t_{on}+1$ stands for the type of the fault,  and $z_{in}=t_{on}+2$ stands for the time stamp, respectively.
 For example, suppose the fault at the time stamp $z_{out}+t_{on}$ includes the trip of line 100 and the fault type is line-to-line fault. We encode this fault type as number $20$ and the fault location as number $100$. Then  $\left(\bm{g}_1\right)_{z_{out}, x,y,t_{on}}=100$, $\left(\bm{g}_1\right)_{z_{out}, x,y,t_{on}+1}=20$ and $\left(\bm{g}_1\right)_{z_{out}, x,y,t_{on}+2}=(z_{out}+t_{on})\Delta t$  for all $x$ and $y$. Namely, the fault type as number $20$, the fault location as number $100$, and the time  $(z_{out}+t_{on})\Delta t$  are duplicated  along the dimension of $x$ and $y$. This way, fault-clearing actions and the output
time stamps are encoded in the dataframe without adding extra
complexity for batch operation.}

After such an encoder, the data frame is converted from a 3D tensor with dimension $\mathbb{R}^{N\times Y \times \tau_{in}}$ to a 4D tensor with dimension $\mathbb{R}^{\tau_{out}\times N\times Y \times (\tau_{in}+3)}$.
\highlights{We train the neural network on a batch of trajectories and the number of trajectories is $B$. Hence, the full size of input data is with the dimension $\mathbb{R}^{B\times\tau_{out}\times N\times Y \times (\tau_{in}+3)}$. We use the  Great Britain transmission network with 2224 nodes that will appear later in the case study to give an impression of the size of the data. For the time steps $\tau_{in}=20$ and $\tau_{out}=150$, the memory space for one input trajectory ($B$=1) in the   Great Britain transmission network is $0.016$GB, and the memory space for 200 input trajectory ($B$=200) in training is $3.2$GB. The memory space for other sizes of systems will scale linearly with the number of buses.}

The faults of interest in the paper are mainly transmission line faults, and therefore we envision that the $\bm{p}$ and $\bm{q}$ during  the prediction horizon of less then $10$s will not deviate too much from the values in the input trajectory. Hence, the encoding of  $\bm{p}$ and $\bm{q}$ in  horizon of the input trajectory already provides enough information to guide a good prediction.
The goal is to use neural networks to map the the input tensor to output tensor  with dimension $\mathbb{R}^{\tau_{out}\times N\times 3}$ for the predicted dynamics of $\delta$, $\omega$ and $V$ along $\tau_{out}$ time steps for $N$ buses. 

\subsection{3D Fourier Transform}
After the encoder, the Fourier transform in~\eqref{eq: Fourier_operator} is reduced to 3D Fourier transform along the axis of $x$, $y$ and $z_{out}$ computed as
\begin{equation}\label{eq:Fourier}
\color{black}
\begin{split}
&(\mathcal{F} \bm{g}_{j})_{\xi_{1}, \xi_{2}, \xi_{3}, z_{in}}
\\
=&\sum_{z_{out}=0}^{\tau_{out}-1}\sum_{x=0}^{N-1}\sum_{y=0}^{Y-1} e^{-2 \pi i\left(\frac{z_{out}\xi_{1}}{\tau_{out}}+\frac{x \xi_{2}}{N}+\frac{y \xi_{3}}{Y} \right)}
 \cdot (\bm{g}_{j})_{z_{out},x, y, z_{in}},
\end{split}
\end{equation}
where $\xi_1$, $\xi_2$ and $\xi_3$ are modes in the frequency domain in the three dimensions after the discrete Fourier transform. \highlight{ Importantly, 3D FFT has been supported by most machine learning frameworks (e.g., Pytorch), which is computational efficient for both backward propagation in training and forward propagation in prediction.}

The structure of Fourier layer with the 3D Fourier Transform is visualized in Fig.~\ref{fig:FNO_structure}(c). After truncating the Fourier series at a maximum number of modes $k_{max,i}$ for the $i$-th dimension, an equivalent convolution in the frequency domain is conducted using dot-product with weights $\bm{\theta}_j\in \mathbb{R}^{k_{max,1}\times k_{max,2}\times k_{max,3} \times (\tau_{in}+3)\times (\tau_{in}+3)}$ defined by
\begin{equation}\label{eq:tensor_multi}
\color{black}
    \left(\bm{\theta}_j \cdot\psi \left(\mathcal{F} \bm{g}_{j}\right)\right)_{\xi_1,\xi_2,\xi_3,  z_{in}}=\sum_{v=0}^{\tau_{in}+2} (\bm{\theta}_j)_{\xi_1,\xi_2,\xi_3,  z_{in}, v} \left(\mathcal{F} \bm{g}_{j}\right)_{\xi_1,\xi_2,\xi_3, v}
\end{equation}
\highlight{for $\xi_1=0, \ldots, k_{\max,1 }-1$, $\xi_2=0, \ldots, k_{\max,2 }-1$, $\xi_3=0, \ldots, k_{\max,3 }-1$, $z_{in}=0, \ldots, \tau_{in}+2$ and $j=1,\ldots, l$.}

\highlight{The time domain signal is recovered by inverse Fourier transform as follows:}
\begin{equation}\label{eq:Inv_Fourier_3D}
\color{black}
\begin{split}
&\left(\mathcal{K}_{\bm{\theta}_j}\left( \bm{g}_{j}\right)\right)_{z_{out},x, y, z_{in}}
\\
=&\left({\mathcal{F}}^{-1}\left(\bm{\theta}_j \cdot\psi \left(\mathcal{F} \bm{g}_{j}\right)\right)\right)_{z_{out},x, y, z_{in}}
\\
=&\sum_{\xi_1=0}^{k_{\max,1 }-1}\sum_{\xi_2=0}^{k_{\max,2 }-1}\sum_{\xi_3=0}^{k_{\max,3 }-1} e^{2 \pi i\left(\frac{z_{out}\xi_{1}}{\tau_{out}}+\frac{x \xi_{2}}{N}+\frac{y \xi_{3}}{Y} \right)}\\
& \quad
 \cdot \left(\bm{\theta}_j \cdot\psi \left(\mathcal{F} \bm{g}_{j}\right)\right)_{\xi_1,\xi_2,\xi_3,  z_{in}},
\end{split}
\end{equation}
\highlight{for $z_{out}=0, \ldots, \tau_{out}-1$, $x=0, \ldots, N-1$, $y=0, \ldots, Y-1$,  $z_{in}=0, \ldots, \tau_{in}+2$ and $j=1,\ldots, l$.}

\highlight{Plugging~\eqref{eq:Inv_Fourier_3D} into~\eqref{eq: Fourier_layer} gives the output of the $j$-th layer $\bm{g}_{j+1}\in \mathbb{R}^{\tau_{out}\times N\times Y \times (\tau_{in}+3)}$.} The predicted trajectory $\hat{\bm{s}}_{out}\in \mathbb{R}^{\tau_{out}\times N\times 3}$ are obtained from the output of the last layer after a dense combination weight $\bm{W}_{l+1 }\in\mathbb{R}^{(\tau_{in}+3)\times 1}$. 
\begin{equation}\label{eq: Fourier_predict_3D}
\color{black}
\left(\hat{\bm{s}}_{out}\right)_{z_{out},x,y}=\sum_{z_{in}=0}^{\tau_{in}+2}
\left(\bm{g}_{l+1}\right)_{z_{out},x,y,z_{in}}\left(\bm{W}_{l+1 }\right)_{z_{in}}
\end{equation}
\highlight{for $z_{out}=0, \ldots, \tau_{out}-1$, $x=0, \ldots, N-1$, $y=0,1, 2$. The index $y=0,1, 2$ corresponds to the prediction of angle, angle speed, and voltage, respectively. Typically, increased layer $l$ enable the structure to learn more complex dynamic patterns.  In simulation, we found that four layers are sufficient. In practice, the encoder may also conduct a linear combination on the 4D tensor along the dimension of $z_{in}$ to increase the representation capability of the neural networks~\cite{li2020fourier}. In that case, the dimension of $\bm{W}_{l+1 }$ needs to be adjusted accordingly and all the other computations still remain the same.  }

\vspace{-0.2cm}
\highlight{\subsection{Algorithm}}
\highlight{
 The pseudo-code for our proposed method is given in Algorithm 1. The variables to be trained are weights $\bm{\Phi}
=\{\bm{\theta},\bm{W}\}$ shown in Fig.~\ref{fig:FNO_structure_total}. 
Adam algorithm is adopted to update weights in each episode.}
\highlights{The main practical benefit is that the learned neural networks are feedforward functions, which can be evaluated orders-of-magnitude faster than conventional iterative solvers. To be clear, these neural networks are not replacement for conventional solvers. Rather, they can be used by system operators to study a much larger set of scenarios of how the system would behave under various types of disturbances. They would be valuable tools that would enable better characterization of the dynamic behavior of the systems, and complement existing tools such as high fidelity simulators.}
\\
\begin{algorithm}
 \caption{\color{black}Training and Predicting Power System Transients}
 \begin{algorithmic}[1]
 \renewcommand{\algorithmicrequire}{\textbf{Training: }}
 \renewcommand{\algorithmicensure}{\textbf{Predicting:}}
 \REQUIRE Learning rate $\alpha$, batch size $H$, number of episodes $I$, dataset for training\\
\textit{Initialisation} :Initial weights $\bm{\Phi}$ of the neural network
  \FOR {$episode = 1$ to $I$}
  \STATE Encode the input trajectory $\bm{s}_{in}^{i}$ and the output trajectory $\bm{s}_{out}^{i}$  for  $i=1,\cdots,H$ with fault information in the dataset
  \STATE Using the current weights $\bm{\Phi}$ of the neural network,  compute the predicted trajectory $\hat{\bm{s}}_{out}^{i}$  for $i=1,\cdots,H$\\
  \STATE  Calculate total loss of all the batches $Loss=\frac{1}{H}\sum_{i=1}^{H}\frac{||\hat{\bm{s}}_{out}^i-\bm{s}_{out}^{i}||_1}{||\bm{s}_{out}^{i}||_1}.$\\
  \STATE  Update weights in the neural network by passing $Loss$ to Adam optimizer:
  $\bm{\Phi} \leftarrow \bm{\Phi}-\alpha \text{Adam}(Loss)$
  \ENDFOR
  
   \ENSURE Pre-trained weights $\bm{\Phi}$ of the neural network
   \STATE Encode the input trajectory $\bm{s}_{in}$ with the setup of fault-clear actions
  \STATE Using the pre-trained weights $\bm{\Phi}$ of the neural network, compute the predicted trajectory $\hat{\bm{s}}_{out}$\\
 \end{algorithmic} 
 \end{algorithm}

    


\section{Case Study} \label{sec:simulation}

    \begin{figure}[ht]	
	\centering
	\includegraphics[width=3in]{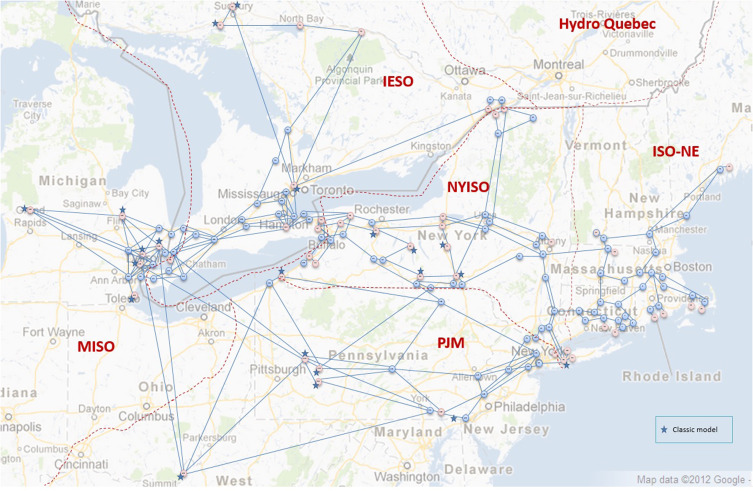}
	\caption{Topology of Northeastern Power Coordinating Council (NPCC) 48-machine,
140-bus power system~\cite{chow1992toolbox,ju2015simulation}}
	\label{fig:NPCC}
    \end{figure}
    
In this section, we conduct several case studies to illustrate the effectiveness of the proposed method. We validate the performance of the proposed approach on a realistic power grid by the case studies with the Northeastern Power Coordinating Council (NPCC) 48-machine, 140-bus test system as shown in Fig.~\ref{fig:NPCC}~\cite{chow1992toolbox,ju2015simulation}. \highlight{Last, we verify the performance on large-scale power systems using  the Great Britain (GB) transmission network, which consists of 2224 nodes, 3207 branches and 394 generators~\cite{cui2020hybrid}. } 
In the appendix\ref{app: single}, a
 simple single-machine infinite bus system is used to show the benefit of learning in the frequency domain compared to the time domain.


\subsection{Simulation and Hyper-Parameter Setup}\label{subsec:Simulation Setting} 
We construct the network in Fig.~\ref{fig:FNO_structure_total}  with four Fourier layers. \highlight{We  normalize the data of different physical meanings and thus eliminate the effect of the magnitude of features brought by different  unit.} The maximum number of modes in the frequency domain is set to be $k_{max,1}=6$, $k_{max,2}=3$ and $k_{max,3}=3$. The episode number and the batch size are 4000 and 800, respectively. Weights of neural networks are updated using Adam with learning rate initializes at 0.02 and decays every 100 steps with a base of 0.85. 
We use Pytorch and a single Nvidia Tesla P100 GPU with 16GB memory. \highlight{We use generic deep neural network (DNN) as a benchmark to compare the performance.
The DNN has a dense structure and seven layers with ReLU activation, where the width of each layer is 20. The hyper-parameters of DNN are also tuned to achieve their best performances for different tasks. The episode number and the batch size are set the same as FNO. }

\begin{figure*}[ht]
\centering
\subfloat[True trajectories (lines) and envelope (grey area) after a three-phase line fault between bus 54 and bus 103 and recovered at the time of 0.3s ]{\includegraphics[width=0.7\textwidth]{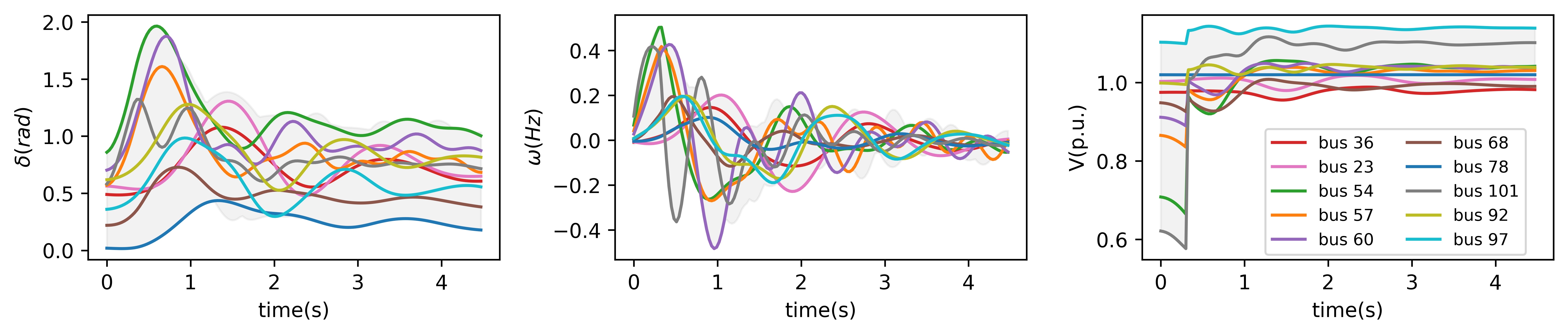}%
}
\vfil
\subfloat[Predicted trajectories (lines) and envelope (grey area) after a three-phase line fault between bus 54 and bus 103  and recovered at the time of 0.3s ]{\includegraphics[width=0.7\textwidth]{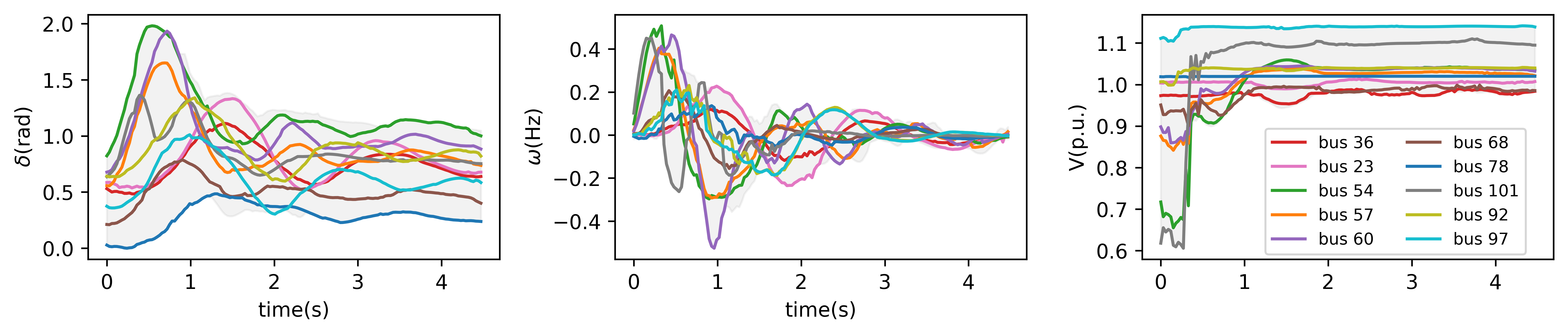}%
}
\caption{Stable dynamics of angle $\delta$ (left), frequency deviation $\omega$ (middle)  and voltage $V$ (right) in NPCC corresponding to (a)  the ground truth produced by a solver. (b) prediction of FNO. The grey area shows the envelope of the trajectories for all generator buses. Lines with different colors shows the trajectories in selected generator buses. The proposed method predict both the magnitude and oscillations accurately.}
\label{fig:dyn_stable}
\end{figure*}

\begin{figure*}[ht]
\centering
\subfloat[True trajectories (lines) and envelope (grey area) after a  line-to-line fault between bus 75 and bus 124 and recovered at the time of 0.3s ]{\includegraphics[width=0.7\textwidth]{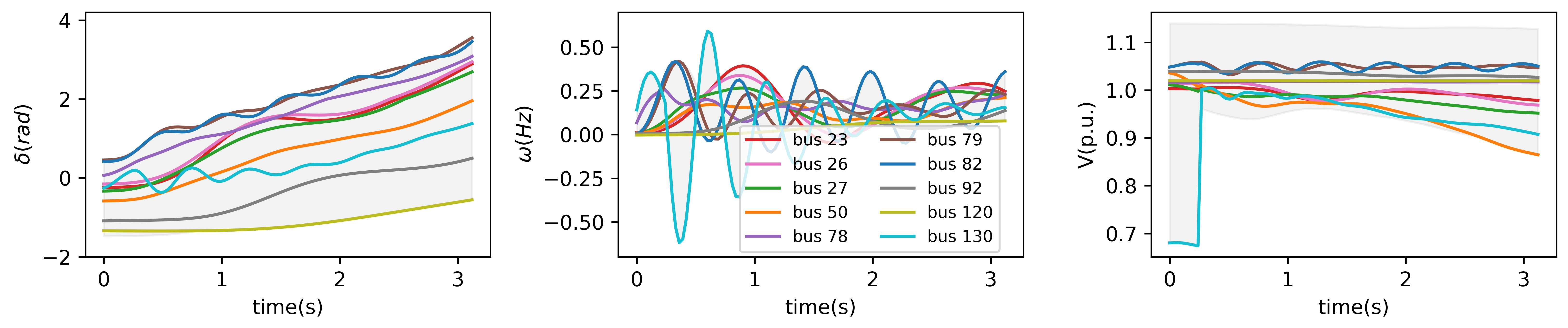}%
}
\vfil
\subfloat[Predicted trajectories (lines) and envelope (grey area) after  a line-to-line fault between bus 75 and bus 124 and recovered at the time of 0.3s ]{\includegraphics[width=0.7\textwidth]{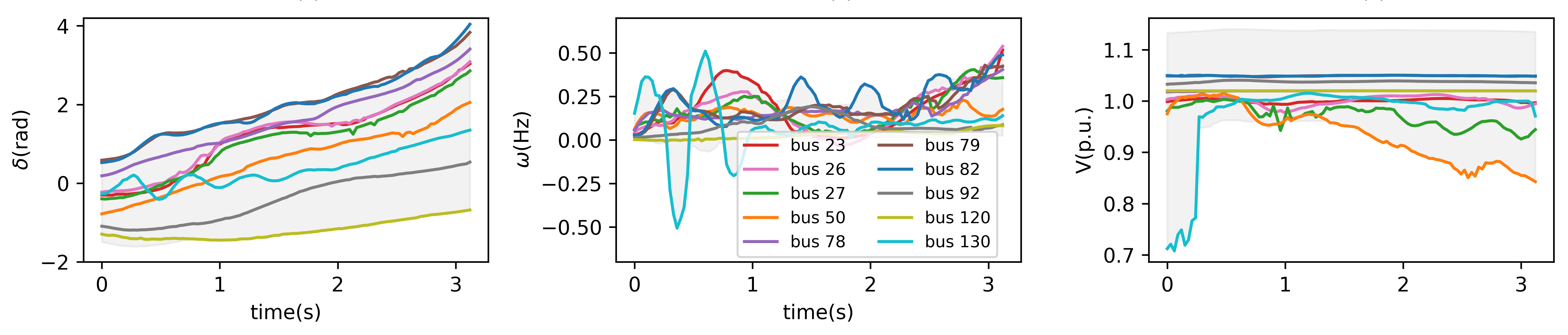}%
}
\caption{Unstable dynamics of angle $\delta$ (left), frequency deviation $\omega$ (middle)  and voltage $V$ (right) in NPCC corresponding to (a)  the ground truth produced by a solver. (b) prediction of FNO. The grey area shows the envelope of the trajectories for all generator buses. Lines with different colors shows the trajectories in selected generator buses. The proposed method predict both the magnitude and oscillations accurately.}
\label{fig:dyn_unstable}
\end{figure*}

\begin{figure*}[ht]
\centering
\subfloat[FFT of true trajectories (lines) and envelope (grey area) after a three-phase line fault between bus 54 and bus 103 and recovered at the time of 0.3s ]{\includegraphics[width=0.7\textwidth]{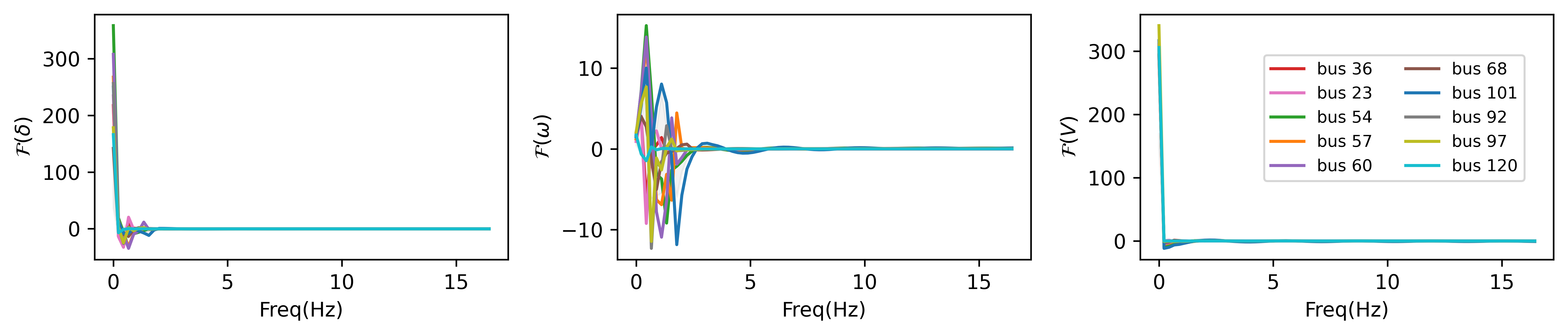}%
}
\vfil
\subfloat[FFT of predicted  trajectories (lines) and envelope (grey area) after a three-phase line fault between bus 54 and bus 103  and recovered at the time of 0.3s ]{\includegraphics[width=0.7\textwidth]{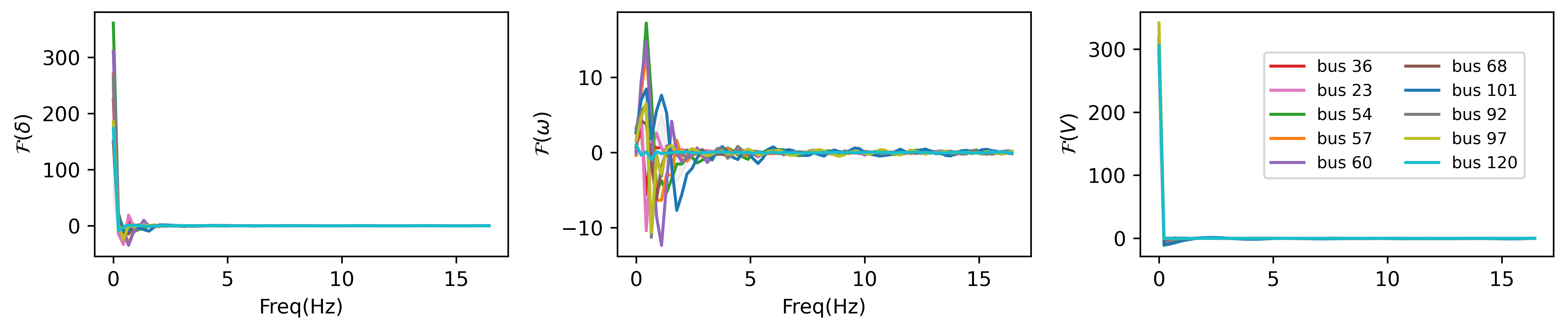}%
}
\caption{Frequency-domain trajectories of angle $\delta$ (left), angle speed deviation $\omega$ (middle)  and voltage $V$ (right) for stable dynamics in NPCC corresponding to (a)  the ground truth produced by a solver. (b) prediction of FNO. }
\label{fig:FFT_dyn_stable}
\end{figure*}

\begin{figure*}[ht]
\centering
\subfloat[FFT of true trajectories (lines) and envelope (grey area) after a  line-to-line fault between bus 75 and bus 124 and recovered at the time of 0.3s ]{\includegraphics[width=0.7\textwidth]{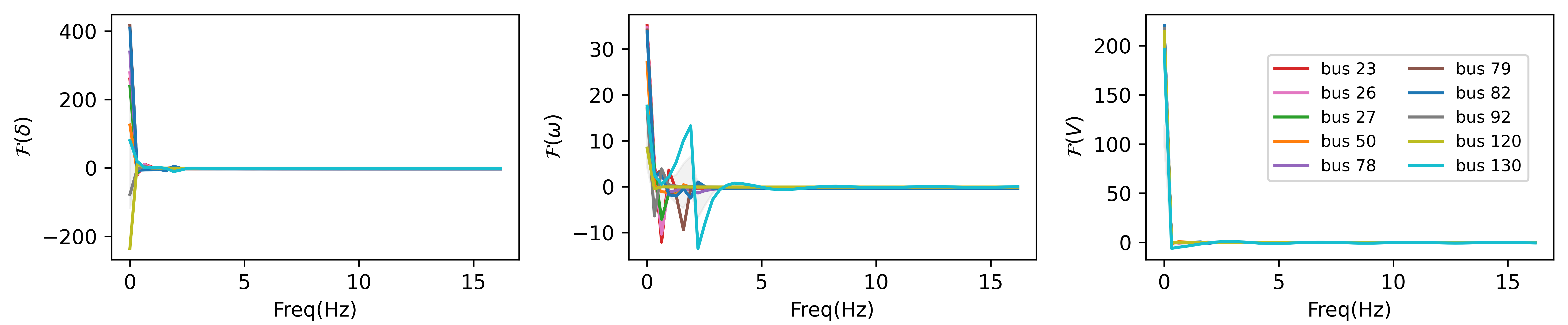}%
}
\vfil
\subfloat[FFT of predicted trajectories (lines) and envelope (grey area) after  a  line-to-line fault between bus 75 and bus 124 and recovered at the time of 0.3s ]{\includegraphics[width=0.7\textwidth]{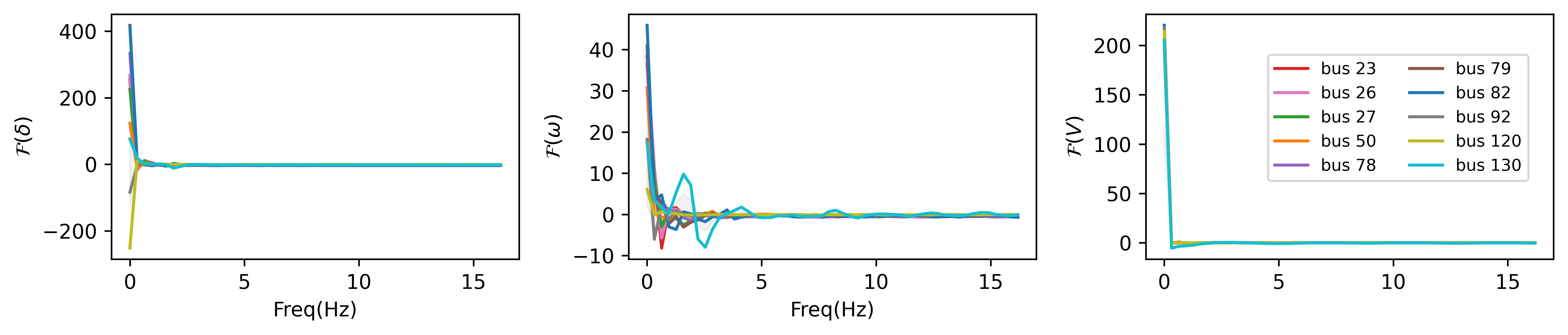}%
}
\caption{Frequency-domain trajectories of angle $\delta$ (left), angle speed deviation $\omega$ (middle)  and voltage $V$ (right) for unstable dynamics in NPCC corresponding to (a) true trajectories the ground truth produced by a solver. (b) prediction of FNO.  }
\label{fig:FFT_dyn_unstable}
\end{figure*}

\subsection{Performance on NPCC test system}
The performance of the proposed method on a practical power system is verified by simulations on  Northeastern Power Coordinating Council (NPCC) test system, which represents the power grid of the northeastern United States and Canada and was involved in the
2003 blackout event~\cite{ju2015simulation}. The power system toolbox in MATLAB is used to generate dataset of power system transient dynamics with the full 6-order generator model, turbine-governing system and exciters~\cite{chow1992toolbox}. \highlight{The power system toolbox utilizes kron-reduced admittance matrix and simulate  dynamics with equivalent ordinary differential equations~\cite{chow1992toolbox}. }
\highlight{ The trajectories are generated considering the actions of protective relays in 4-20 cycles~\cite{anderson2022power}. We create cases  with stressed conditions (stable and unstable) by increasing the level of loads until the system is unstable. The cases with stressed conditions account for  15\% in the training set. }

The input trajectories evolves $\tau_{in} = 20$ time steps, with time interval $\Delta t$ between neighbouring time step to be 0.03 (i.e., approximate two cycles that can be attained by most phasor measurement unit (PMU)). We predict the subsequent trajectories of the length $\tau_{out} = 150$ for total duration of \highlight{$4.5$s}. 
\highlight{The training time of 4000 episodes is 4424.33s.  We quantify the prediction accuracy through the relative mean squared error (RMSE) defined as 
$||\bm{s}_{out}-\hat{\bm{s}}_{out}||_2^2/||\bm{s}_{out}||_2^2$.}

Fig.~\ref{fig:dyn_stable} shows true (i.e., simulated)  and predicted trajectories of the system after  a three-phase line fault between bus 54 and bus 103  cleared at the time of 0.3s. Let the time of fault happens as the time $t=0$. The prediction starts at $t_{on}=0.06s$, which means one time step in the input trajectories corresponds to the fault-on system. The grey area is the envelope of trajectories in all generator buses and the lines are the trajectories in ten generator buses. For all the three states variables (i.e., $\delta$, $\omega$ and $V$), the predicted trajectories in Fig.~\ref{fig:dyn_stable}(b) has similar envelope as the accurate trajectories in  Fig.~\ref{fig:dyn_stable}(a). \highlight{The RMSE for the prediction in Fig.~\ref{fig:dyn_stable} is 0.0041.} The convergence of the envelope in frequency deviation $\omega$ to zeros indicates that the system is stable after the fault and its clear action. Moreover, both the magnitude and the periodic oscillations in the ten generator buses are all captured by the prediction with FNO for the on-fault and post-fault period. As illustrated in Fig.\ref{fig:FNO_structure}, the type of fault and the fault-clear action at $t=0.3s$ is encoded in the input tensor of FNO. Correspondingly, Fig.~\ref{fig:dyn_stable}(b) predicts a step increase of voltage at  $t=0.3s$, which is the same as  Fig.~\ref{fig:dyn_stable}(a). Notably, the magnitude of voltage at the bus 54 and bus 101 is below 0.8 p.u. before $t=1s$, exceeding the permissible  ranges of 5\% from nominal. This may cause low-voltage curtailment of the generators and warrant attention from system operators. Therefore, the proposed prediction can provide sufficient information for identifying how danger the system is.

To illustrate the performance of the proposed method in predicting an unstable system, \highlight{Fig.~\ref{fig:dyn_unstable} shows true (i.e., simulated) and predicted trajectories of the system after a line-to-line fault between bus 75 and bus 124 and recovered at the time of 0.3s. Especially, it is a stressed unstable test case by gradually increasing the level of loads until the system is unstable. Although the magnitude of the trajectories are still bounded, the drifted angle and the collapse of voltage have reflected the unstable behaviors.  The proposed method capture both the trend and oscillations of the unstable behaviors.} 
\highlight{The RMSE for the prediction in Fig.~\ref{fig:dyn_unstable} is 0.1198. Moreover, the 
accuracy in terms of predicting unstable cases is to identify the unstable behaviors. In the next subsection, we verify in the test set that the proposed method can predict all the unstable system accurately shortly after a fault happens. 
}

\highlight{The trajectory in the frequency domain (computed by Fast Fourier Transform) for the stable case in Fig.~\ref{fig:dyn_stable} and the unstable case in Fig.~\ref{fig:dyn_unstable} is given in Fig.~\ref{fig:FFT_dyn_stable} and Fig.~\ref{fig:FFT_dyn_unstable}, respectively. The proposed method also achieves high accuracy in the frequency domain. Moreover, the high-frequency component in both Fig.~\ref{fig:FFT_dyn_stable} and Fig.~\ref{fig:FFT_dyn_unstable} is almost zero. This provides the intuition why the low-pass filter can reduce computational complexity without affecting the prediction performances. }

\highlights{Numerical studies about the effect of $k_{\max}$ on the low-pass filter can be found in Appendix\ref{app: filter}.
The visualization of the predictions with different influence factors including fault-on/clear actions, fault type and fault location is shown in Appendix\ref{app: factor}.}

\subsection{Quantifying the Performance on NPCC}
As shown in Fig.~\ref{fig:dyn_stable} and Fig.~\ref{fig:dyn_unstable}, the dynamics of the power system transient states differ greatly with different fault types and system parameters. To quantify the performance of the proposed method in  stochastic scenarios, we calculate the mean prediction error in the test set with 100 cases where initial states, location of fault, type of fault and fault-clearing time are randomly generated. Three metrics are included:
\begin{enumerate}
    \item Relative mean squared error (RMSE)
    
    \item Type1-error: Percentage of unstable cases predicted to be stable. This is the more severe type of error and may cause blackouts of power systems (instability is declared when the average value of $\omega$ from $t=4s$ to $t=4.5s$ exceed  0.5Hz).   
    \item Type2-error: Percentage of stable cases predicted to be unstable. 
\end{enumerate}

\begin{table*}[ht]
\renewcommand{\arraystretch}{1.3}
\caption{Performance - on fault }
\label{tab:metric_on_fault}
\begin{tabular}{c||c|c|c|c|c||c|c|c|c|c||c|c|c|c|c}
\hline \hline
Metric &  \multicolumn{5}{c||}{Relative mse}  &  \multicolumn{5}{c||}{Error-Type1}  &  \multicolumn{5}{c}{Error-Type2}\\
\hline
Cycle after fault & 0 & 2 & 4 & 10 & 20
& 0 & 2 & 4  & 10 & 20
& 0 & 2 & 4 & 10 & 20 \\
\hline
FNO & $\bm{0.0144}$ & $\bm{0.0063}$ & $\bm{0.0055}$ & $\bm{0.0053}$ & $\bm{0.0051}$
& $\bm{0}$ & $\bm{0}$ & $\bm{0}$ & $\bm{0}$ & $\bm{0}$
 & 0.011 & 0 & 0 & 0 & 0\\
\hline
DNN & 0.0778  & 0.0712 & 0.0655& 0.0654  & 0.0656 
& 1 & 1 & 1 & 0.667   & 0.167 
& 0 & 0 & 0 & 0.011 & 0 \\
\hline\hline
\end{tabular}
\end{table*}

Notably, we fix the length of the input  trajectories $\bm{s}_{in}$ to be $\tau_{in} = 20$. \highlight{The input trajectories contain data before and after the fault (similar to a rolling window). As illustrated in Fig. \ref{fig:prediction}, more data point after fault will be observed if the prediction starting point $t_{on}$ is larger (if we wait longer after the fault to do a prediction). The more steps after the fault in $\bm{s}_{in}$, the better the prediction performance. }
Table~\ref{tab:metric_on_fault}  summarize the metrics for the prediction error corresponding to different number of on-fault cycles (one cycle is  1/60=0.017s) involved in the input trajectories $\bm{s}_{in}$.


\highlight{From Table~\ref{tab:metric_on_fault}, 
the RMSE of FNO is much lower than the case in DNN, respectively.}
Interestingly, DNN has the Type2-error to be approximate zero while extremely high Type1-error. The reason is that DNN will easily overfit since the majority (93\%) of training samples. By contrast, FNO brings zero Type1-error and Type2-error once there is an on-fault data point entered in the input trajectories. Therefore, the proposed prediction with FNO will capture all the dangerous unstable case.
The low RMSE indicates that the proposed method can also simulate the dynamics of trajectories accurately.


\highlight{Considering that unstable behaviors may not be observed in real measurements,  we also investigate the performance of the proposed method where unstable behaviors are not present in the training set. Table~\ref{tab:metric_wo_stressed} shows the comparison of RMSE on the dataset without unstable cases. The RMSE decreases greatly after eliminating the unstable cases. Of course, this is an easier problem, and the RMSE decreases greatly after eliminating the unstable cases.}

\begin{table}[ht]
\renewcommand{\arraystretch}{1.3}
\caption{RMSE for the dataset without stressed unstable cases}
\label{tab:metric_wo_stressed}
\centering
\begin{tabular}{c||c|c|c|c|c}
\hline \hline
Cycle after fault & 0 & 2 & 4 
& 10 & 20 \\
\hline
FNO & 0.0077 & 0.0011 &  0.0009
& 0.0008 &  0.0007
\\
\hline
DNN 
& 0.0154 &  0.0146 &  0.0141 
&  0.0125 &  0.0104
 \\
\hline\hline
\end{tabular}
\end{table}

\begin{table}[!h]
\renewcommand{\arraystretch}{1.3}
\caption{Average Computational Time }
\label{tab:Computational_time}
\centering
\begin{tabular}{cccc}
\hline \hline
Methods & FNO & Matlab toolbox & Speed up\\
\hline
Prediction horizon = 3s & 0.0036 & 1.69 & 469x\\
\hline
Prediction horizon = 4.5s & 0.0037& 2.24 & 605x \\
\hline
Prediction horizon = 6s & 0.0039 &  3.38 & 867x \\
\hline\hline
\end{tabular}
\end{table}
Lastly, we compare the average  computational time in the test set for  FNO and Power System Toolbox in MATLAB as shown in Table~\ref{tab:Computational_time}. \highlight{ The execution time for 3D Fourier Transform and 3D inverse Fourier Transform in one layer is $7.79\times 10^{-5}$s and $1.21\times 10^{-4}$s, respectively.}
For the prediction time horizon ranges from 3s, 4.5, and 6s, the computational time of FNO is  0.0036s, 0.0037s, 0.0039s, respectively. By constrast, the computational time of MATLAB toolbox are 469, 605 and 867 times slower than FNO. Therefore, the proposed approach will significantly speed up the simulation for power system transient dynamics.



\subsection{Case study on the Great Britain  transmission network}

We conduct case study on GB system to test the performance of the proposed method on large power networks.  We use ANDES (an open source package for power system dynamic simulation)  to  generate dataset of power system transient dynamics with the full 6-order generator model, turbine-governing system and exciters~\cite{cui2020hybrid}. Differential-algebraic equations are solved for dynamic simulation~\cite{cui2020hybrid}. The input trajectories evolves $\tau_{in} = 20$ time steps, with a sampling period of 1/30s (i.e., two cycles). We predict the subsequent trajectories of the length $\tau_{out} = 150$ steps, for a total duration of 5s. \highlights{The number of trajectories we used to train the neural network is 200.}  The training time of 4000 episodes  is 9256.96s. \highlights{We believe that increasing the number of trajectories can further improve the performance of the learned neural networks, but our computing resources (a single Nvidia Tesla P100 GPU with 16GB memory) limit us to 200 trajectories. Despite of this, the following simulation result on the test set shows that the performance of the learned neural network is actually good enough. }

\begin{figure*}[ht]
\centering
\subfloat[True trajectories (lines) and envelope (grey area) after a three-phase-to-ground fault between bus 56 and bus 637 and recovered at the time of 0.13s ]{\includegraphics[width=0.7\textwidth]{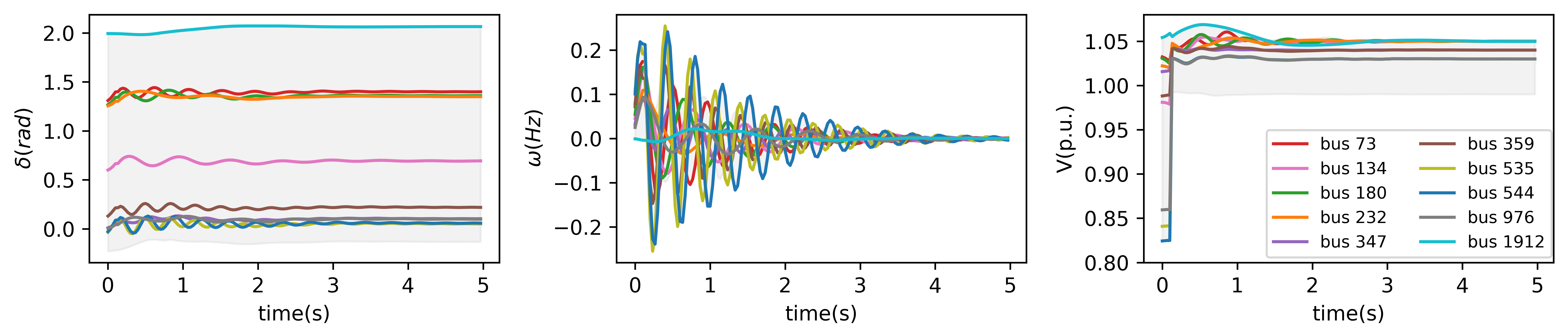}%
}
\vfil
\subfloat[Predicted trajectories (lines) and envelope (grey area) after a three-phase-to-ground fault between bus  56 and bus 637 and recovered at the time of 0.13s ]{\includegraphics[width=0.7\textwidth]{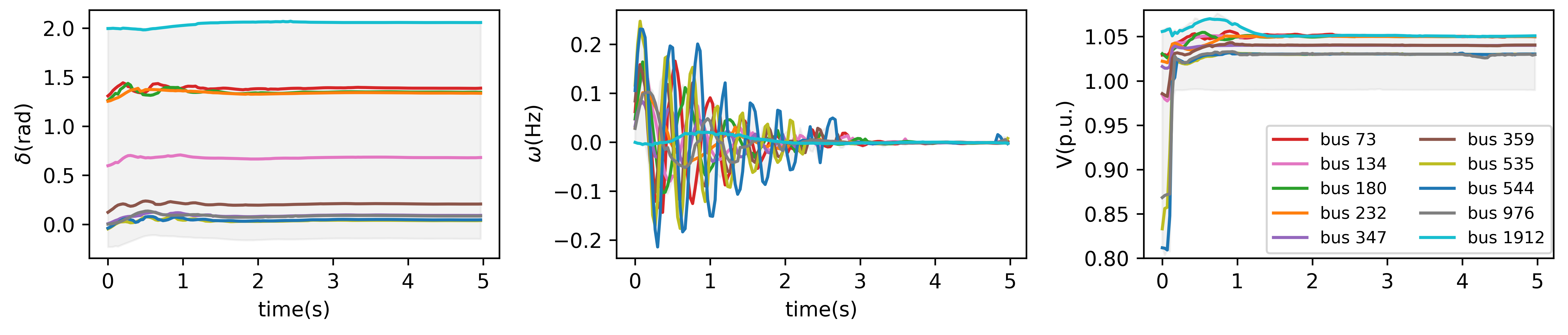}%
}
\caption{Dynamics of angle $\delta$ (left), frequency deviation $\omega$ (middle)  and voltage $V$ (right) in GB corresponding to (a) the ground truth produced by a solver (b) prediction of FNO. The grey area shows the envelope of the trajectories for all generator buses. Lines with different colors shows the trajectories in selected generator buses. The proposed method predict both the magnitude and oscillations accurately.}
\label{fig:dyn_stable_GB}
\end{figure*}

 Let the fault happens as the time $t=0$. The prediction starts at $t_{on}=2/30s$, which means one time step in the input trajectories corresponds to the on-fault system. \highlights{ Fig.~\ref{fig:dyn_stable_GB}  shows the prediction on a line fault (a three-phase-to-ground fault between bus  56 and bus 637 and recovered at the time of 0.13s) that is not covered in the training set. }
 The grey area is the envelope of trajectories in all generator buses and the lines are the trajectories in ten generator buses. For all the three states variables (i.e., $\delta$, $\omega$ and $V$), the predicted trajectories in Fig.~\ref{fig:dyn_stable_GB}(b) have similar envelope as the simulated (i.e., ground-truth) trajectories in  Fig.~\ref{fig:dyn_stable_GB}(a).  Moreover, both the magnitude and the periodic oscillations in the ten generator buses are all captured by the prediction with FNO for the on-fault and post-fault period. 

\begin{figure*}[ht]
\centering
\subfloat[True trajectories (lines) and envelope (grey area) after a three-phase-to-ground fault between bus 482 and bus 944 and recovered at the time of 0.13s ]{\includegraphics[width=0.7\textwidth]{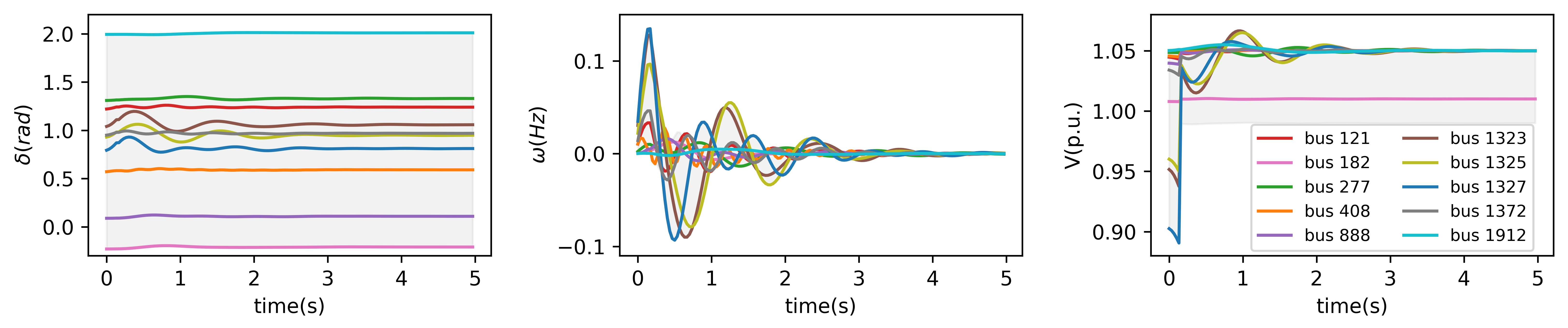}%
}
\vfil
\subfloat[Predicted trajectories (lines) and envelope (grey area) after a three-phase-to-ground fault between bus 482 and bus 944 and recovered at the time of 0.13s ]{\includegraphics[width=0.7\textwidth]{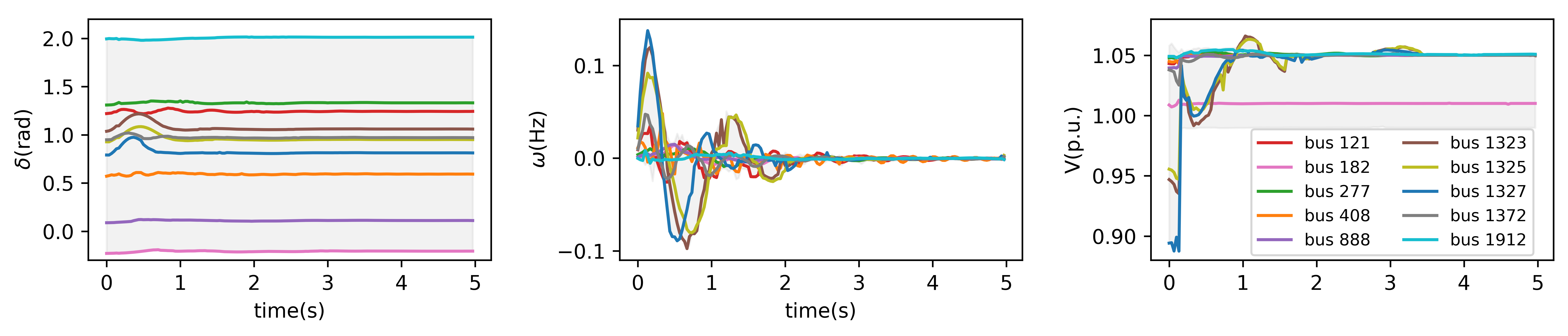}%
}
\caption{\highlights{Dynamics of angle $\delta$ (left), frequency deviation $\omega$ (middle)  and voltage $V$ (right) in GB corresponding to (a) the ground truth produced by a solver (b) prediction of FNO. The proposed method predict both the magnitude and oscillations accurately.}}
\label{fig:dyn_stable_GB_Line1354}
\end{figure*}

 \highlights{Fig.~\ref{fig:dyn_stable_GB_Line1354} shows the prediction on another fault (a three-phase-to-ground fault between bus 482 and bus 944 ) that is far-away from the fault shown in Fig.~\ref{fig:dyn_stable_GB}. The dynamics of Fig.~\ref{fig:dyn_stable_GB} and Fig.~\ref{fig:dyn_stable_GB_Line1354} look to be very different, so the transient under one fault appears to have no relationship
with another far-away fault. Despite of this, Fig.~\ref{fig:dyn_stable_GB} and Fig.~\ref{fig:dyn_stable_GB_Line1354} show that the neural network can capture both patterns of the dynamics. }

\highlight{In  100 test cases where initial states, location of fault, and fault-clearing time are randomly generated, the mean value of RMSE for DNN is 0.0034. By contrast, the mean value of RMSE for
FNO  is 0.0001, which is
97\% smaller than DNN. Hence, the proposed method achieves much higher accuracy compared with generic deep neural networks.  } 

\highlights{ The above experiments show that the neural network is not simply memorizing but generalizing as well~\cite{kawaguchi_bengio_kaelbling_2022,huang2019adaptive}. Since the power system is a synchronized and connected network, transients from different faults could be related. We conjecture that there should be some sparse pattern behind the transient dynamics of the system, and this is the reason why we can learn well with a moderate amount of data. These relationships may be hard to visualize or analytically characterize, which makes machine learning useful. Theoretical analysis of the phenomenon is an important future direction for us. }

\vspace{-0.2cm}
\section{Conclusion} \label{sec:conclusion}
This paper proposes a frequency domain approach for predicting power system transient dynamics. Inspired by the intuition that there are relatively few dominate modes in the frequency domain, we  construct neural  networks  with  Fourier  transform  and  filtering  layers. We design the dataframe to encode the power system topology  and fault-on/clear information in transient dynamics, allowing the extraction of spatial-temporal relationship through 3D Fourier transform. Simulation results show that the proposed approach  speeds  up prediction  computations  by  orders  of  magnitude  and  is  highly accurate  for  different  fault  types. Compared with state-of-the art AI methods, the proposed method reduce MSE prediction error by more than 50\% and vastly improves the detection of unstable dynamics.

\highlights{The simulation results point to an interesting observation that there are sparse patterns behind the transient dynamics of the system, and it is this sparsity that allows the neural networks to learn and predict. Making the theory rigorous is an important future direction for us. The RAM space of GPU resources constrains the amount of  data that can be processed to train the neural networks. Investigating the parallel training on multiple GPU resources and the better usage of RAM space are also important future directions for the proposed method to be utilized in larger systems. }


\bibliographystyle{IEEEtran}
\bibliography{Reference}

\begin{thebibliography}{10}
\providecommand{\url}[1]{#1}
\csname url@samestyle\endcsname
\providecommand{\newblock}{\relax}
\providecommand{\bibinfo}[2]{#2}
\providecommand{\BIBentrySTDinterwordspacing}{\spaceskip=0pt\relax}
\providecommand{\BIBentryALTinterwordstretchfactor}{4}
\providecommand{\BIBentryALTinterwordspacing}{\spaceskip=\fontdimen2\font plus
\BIBentryALTinterwordstretchfactor\fontdimen3\font minus
  \fontdimen4\font\relax}
\providecommand{\BIBforeignlanguage}[2]{{%
\expandafter\ifx\csname l@#1\endcsname\relax
\typeout{** WARNING: IEEEtran.bst: No hyphenation pattern has been}%
\typeout{** loaded for the language `#1'. Using the pattern for}%
\typeout{** the default language instead.}%
\else
\language=\csname l@#1\endcsname
\fi
#2}}
\providecommand{\BIBdecl}{\relax}
\BIBdecl

\bibitem{mai2018electrification}
T.~T. Mai, P.~Jadun, J.~S. Logan, C.~A. McMillan, M.~Muratori, D.~C. Steinberg,
  L.~J. Vimmerstedt, B.~Haley, R.~Jones, and B.~Nelson, ``Electrification
  futures study: Scenarios of electric technology adoption and power
  consumption for the united states,'' National Renewable Energy Lab.(NREL),
  Golden, CO (United States), Tech. Rep., 2018.

\bibitem{golden2015curtailment}
R.~Golden and B.~Paulos, ``Curtailment of renewable energy in california and
  beyond,'' \emph{The Electricity Journal}, vol.~28, no.~6, pp. 36--50, 2015.

\bibitem{kundur2017power}
P.~S. Kundur, N.~J. Balu, and M.~G. Lauby, ``Power system dynamics and
  stability,'' \emph{Power System Stability and Control}, vol.~3, 2017.

\bibitem{sauer2017power}
P.~W. Sauer, M.~A. Pai, and J.~H. Chow, \emph{Power system dynamics and
  stability: with synchrophasor measurement and power system toolbox}.\hskip
  1em plus 0.5em minus 0.4em\relax John Wiley \& Sons, 2017.

\bibitem{chiang2011direct}
H.-D. Chiang, \emph{Direct methods for stability analysis of electric power
  systems: theoretical foundation, BCU methodologies, and applications}.\hskip
  1em plus 0.5em minus 0.4em\relax John Wiley \& Sons, 2011.

\bibitem{florida2008}
{Event Analysis Team}, ``{FRCC} system disturbance and underfrequency load
  shedding event report,'' Operating Committee, Tech. Rep., 2008.

\bibitem{kundur1994power}
P.~Kundur, N.~J. Balu, and M.~G. Lauby, \emph{Power system stability and
  control}.\hskip 1em plus 0.5em minus 0.4em\relax McGraw-hill New York, 1994,
  vol.~7.

\bibitem{huang2017faster}
R.~Huang, S.~Jin, Y.~Chen, R.~Diao, B.~Palmer, Q.~Huang, and Z.~Huang, ``Faster
  than real-time dynamic simulation for large-size power system with detailed
  dynamic models using high-performance computing platform,'' in \emph{2017
  IEEE Power \& Energy Society General Meeting}.\hskip 1em plus 0.5em minus
  0.4em\relax IEEE, 2017, pp. 1--5.

\bibitem{wu2019grid}
Y.-K. Wu, S.-M. Chang, and P.~Mandal, ``Grid-connected wind power plants: A
  survey on the integration requirements in modern grid codes,'' \emph{IEEE
  Transactions on Industry Applications}, vol.~55, no.~6, pp. 5584--5593, 2019.

\bibitem{chalishazar2021power}
V.~H. Chalishazar, S.~Poudel, S.~Hanif, and P.~T. Mana, ``Power system
  resilience metrics augmentation for critical load prioritization,'' PNNL,
  Richland, WA, Tech. Rep., 2021.

\bibitem{ren2022interpretable}
C.~Ren, Y.~Xu, and R.~Zhang, ``An interpretable deep learning method for power
  system transient stability assessment via tree regularization,'' \emph{IEEE
  Transactions on Power Systems}, vol.~37, no.~5, pp. 3359--3369, 2022.

\bibitem{zhu2020intelligent}
L.~Zhu, D.~J. Hill, and C.~Lu, ``Intelligent short-term voltage stability
  assessment via spatial attention rectified rnn learning,'' \emph{IEEE
  Transactions on Industrial Informatics}, vol.~17, no.~10, pp. 7005--7016,
  2020.

\bibitem{zhu2021networked}
L.~Zhu and D.~Hill, ``Networked time series shapelet learning for power system
  transient stability assessment,'' \emph{IEEE Transactions on Power Systems},
  2021.

\bibitem{yan2019fast}
R.~Yan, G.~Geng, Q.~Jiang, and Y.~Li, ``Fast transient stability batch
  assessment using cascaded convolutional neural networks,'' \emph{IEEE
  Transactions on Power Systems}, vol.~34, no.~4, pp. 2802--2813, 2019.

\bibitem{cui2021data}
M.~Cui, F.~Li, H.~Cui, S.~Bu, and D.~Shi, ``Data-driven joint voltage stability
  assessment considering load uncertainty: A variational bayes inference
  integrated with multi-cnns,'' \emph{IEEE Transactions on Power Systems},
  vol.~37, no.~3, pp. 1904--1915, 2021.

\bibitem{xia2018galerkin}
B.~Xia, H.~Wu, Y.~Qiu, B.~Lou, and Y.~Song, ``A galerkin method-based
  polynomial approximation for parametric problems in power system transient
  analysis,'' \emph{IEEE Transactions on Power Systems}, vol.~34, no.~2, pp.
  1620--1629, 2018.

\bibitem{su2021intelligent}
H.-Y. Su and H.-H. Hong, ``An intelligent data-driven learning approach to
  enhance online probabilistic voltage stability margin prediction,''
  \emph{IEEE Transactions on Power Systems}, vol.~36, no.~4, pp. 3790--3793,
  2021.

\bibitem{yang2019pmu}
H.~Yang, W.~Zhang, F.~Shi, J.~Xie, and W.~Ju, ``Pmu-based model-free method for
  transient instability prediction and emergency generator-shedding control,''
  \emph{International Journal of Electrical Power \& Energy Systems}, vol. 105,
  pp. 381--393, 2019.

\bibitem{cepeda2014real}
J.~C. Cepeda, J.~L. Rueda, D.~G. Colom{\'e}, and D.~E. Echeverr{\'\i}a,
  ``Real-time transient stability assessment based on centre-of-inertia
  estimation from phasor measurement unit records,'' \emph{IET Generation,
  Transmission \& Distribution}, vol.~8, no.~8, pp. 1363--1376, 2014.

\bibitem{jalali2022inferring}
M.~Jalali, V.~Kekatos, S.~Bhela, H.~Zhu, and V.~A. Centeno, ``Inferring power
  system dynamics from synchrophasor data using gaussian processes,''
  \emph{IEEE Transactions on Power Systems}, 2022.

\bibitem{stiasny2021learning}
J.~Stiasny, S.~Chevalier, and S.~Chatzivasileiadis, ``Learning without data:
  Physics-informed neural networks for fast time-domain simulation,''
  \emph{arXiv preprint arXiv:2106.15987}, 2021.

\bibitem{li2020fourier}
Z.~Li, N.~Kovachki, K.~Azizzadenesheli, B.~Liu, K.~Bhattacharya, A.~Stuart, and
  A.~Anandkumar, ``Fourier neural operator for parametric partial differential
  equations,'' \emph{arXiv preprint arXiv:2010.08895}, 2020.

\bibitem{jia2021transient}
K.~Jia, Q.~Liu, B.~Yang, L.~Zheng, Y.~Fang, and T.~Bi, ``Transient fault
  current analysis of iiress considering controller saturation,'' \emph{IEEE
  Transactions on Smart Grid}, vol.~13, no.~1, pp. 496--504, 2021.

\bibitem{sauer2021power}
P.~W. Sauer, M.~A. Pai, and J.~H. Chow, \emph{Power system dynamics and
  stability: with synchrophasor measurement and power system toolbox}.\hskip
  1em plus 0.5em minus 0.4em\relax John Wiley \& Sons, 2021.

\bibitem{chow1992toolbox}
J.~H. Chow and K.~W. Cheung, ``A toolbox for power system dynamics and control
  engineering education and research,'' \emph{IEEE transactions on Power
  Systems}, vol.~7, no.~4, pp. 1559--1564, 1992.

\bibitem{raissi2019physics}
M.~Raissi, P.~Perdikaris, and G.~E. Karniadakis, ``Physics-informed neural
  networks: A deep learning framework for solving forward and inverse problems
  involving nonlinear partial differential equations,'' \emph{Journal of
  Computational Physics}, vol. 378, pp. 686--707, 2019.

\bibitem{guo2018graph}
L.~Guo, C.~Zhao, and S.~H. Low, ``Graph laplacian spectrum and primary
  frequency regulation,'' in \emph{2018 IEEE Conference on Decision and Control
  (CDC)}.\hskip 1em plus 0.5em minus 0.4em\relax IEEE, 2018, pp. 158--165.

\bibitem{brunton2019data}
S.~L. Brunton and J.~N. Kutz, \emph{Data-driven science and engineering:
  Machine learning, dynamical systems, and control}.\hskip 1em plus 0.5em minus
  0.4em\relax Cambridge University Press, 2019.

\bibitem{zhao2014design}
C.~Zhao, U.~Topcu, N.~Li, and S.~Low, ``Design and stability of load-side
  primary frequency control in power systems,'' \emph{IEEE Transactions on
  Automatic Control}, vol.~59, no.~5, pp. 1177--1189, 2014.

\bibitem{ju2015simulation}
W.~Ju, J.~Qi, and K.~Sun, ``Simulation and analysis of cascading failures on an
  npcc power system test bed,'' in \emph{2015 IEEE Power \& Energy Society
  General Meeting}.\hskip 1em plus 0.5em minus 0.4em\relax IEEE, 2015, pp.
  1--5.

\bibitem{cui2020hybrid}
H.~Cui, F.~Li, and K.~Tomsovic, ``Hybrid symbolic-numeric framework for power
  system modeling and analysis,'' \emph{IEEE Transactions on Power Systems},
  vol.~36, no.~2, pp. 1373--1384, 2020.

\bibitem{anderson2022power}
P.~M. Anderson, C.~Henville, R.~Rifaat, B.~Johnson, and S.~Meliopoulos,
  \emph{Power system protection}.\hskip 1em plus 0.5em minus 0.4em\relax John
  Wiley \& Sons, 2022.

\bibitem{kawaguchi_bengio_kaelbling_2022}
K.~Kawaguchi, Y.~Bengio, and L.~Kaelbling, \emph{Generalization in Deep
  Learning}.\hskip 1em plus 0.5em minus 0.4em\relax Cambridge University Press,
  2022, pp. 112--148.

\bibitem{huang2019adaptive}
Q.~Huang, R.~Huang, W.~Hao, J.~Tan, R.~Fan, and Z.~Huang, ``Adaptive power
  system emergency control using deep reinforcement learning,'' \emph{IEEE
  Transactions on Smart Grid}, 2019.

\bibitem{ernst2008reinforcement}
D.~Ernst, M.~Glavic, F.~Capitanescu, and L.~Wehenkel, ``Reinforcement learning
  versus model predictive control: a comparison on a power system problem,''
  \emph{IEEE Transactions on Systems, Man, and Cybernetics, Part B
  (Cybernetics)}, vol.~39, no.~2, pp. 517--529, 2008.

\end{thebibliography}
\appendices

\subsection{A Single-Machine Infinite Bus System Example }\label{app: single} 
To visualize and compare the performance of different prediction approaches, we show an illustrative example on a generator connected to an infinite bus, modeling a connection to a large grid that appears as a voltage source~\cite{ernst2008reinforcement}.
The proposed method using Fourier Neural Operator (FNO) is compared with Physics-Informed Neural Networks (PINN) and DNN. The parameters for PINN is the same as~\cite{stiasny2021learning} where the case study is also a single-machine infinite bus system. 
For the prediction with 4.5 seconds, the average relative mse on the test set for FNO, DNN, PINN are 0.0098, 0.1811, and 15.53, respectively. The extreme large mse for PINN reflects that it fails when a long prediction horizon is needed. 

\begin{figure}[ht]	
	\centering
	\includegraphics[width=3.5in]{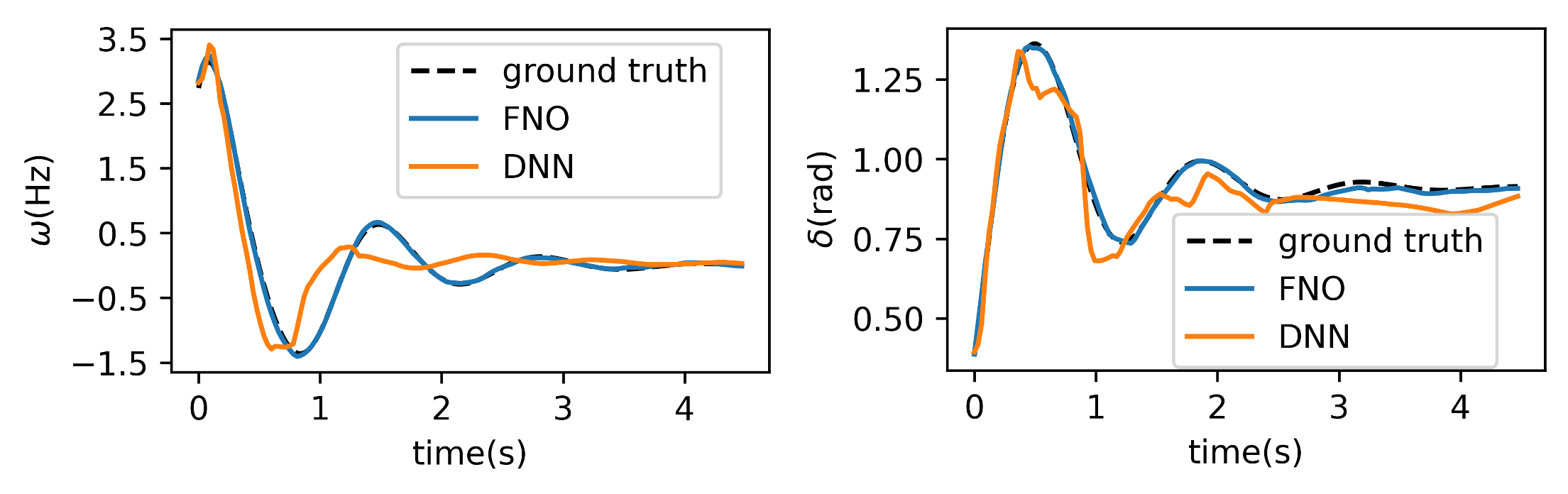}
	\caption{Example trajectories on a single-machine infinite bus system. FNO almost exactly matches the true one, while DNN shows much larger errors. }
	\label{fig:Single_bus_dyn}
    \end{figure}

Fig.~\ref{fig:Single_bus_dyn} shows the dynamics of frequency deviation $\omega$ and angle $\delta$ in a prediction of 4.5 seconds for FNO and DNN. The ground truth is the trajectory found by numerical simulation. FNO fits the ground truth almost perfectly. By contrast, the learned dynamics from DNN are not smooth and show much larger deviations from the ground truth compared with FNO. This verifies the analysis illustrated in Fig.~\ref{fig:Intuition_ML} that purely time-domain tends to overfit easily and have difficulties learning smooth oscillations.
    
\subsection{\highlights{The influence of the low-pass filter}\label{app: filter}}

\begin{figure*}[ht]
\centering
\subfloat[Simulated]{\includegraphics[width=0.3\textwidth]{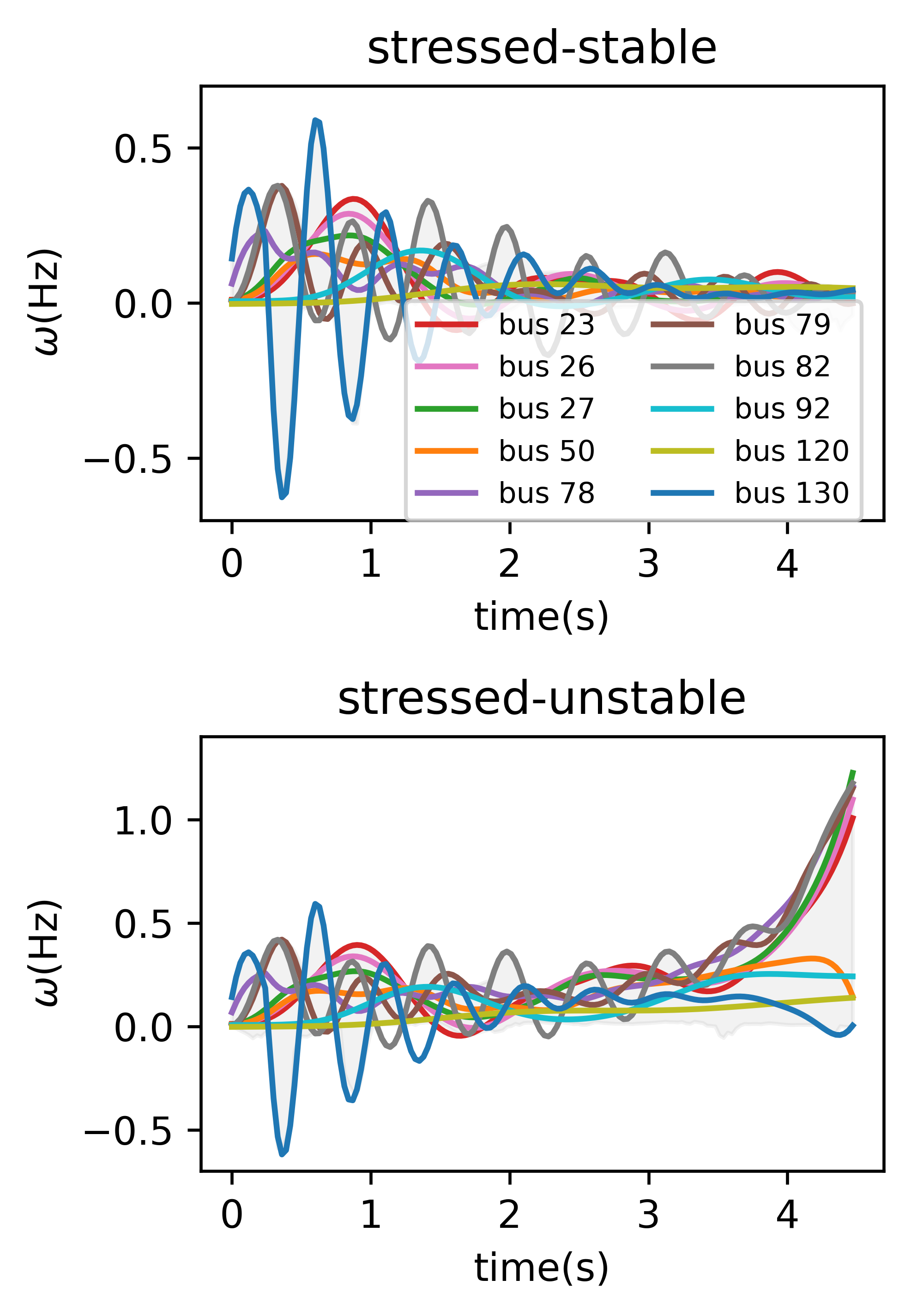}%
}
\hfil
\subfloat[w.o. low-pass filter]
{\includegraphics[width=0.3\textwidth]{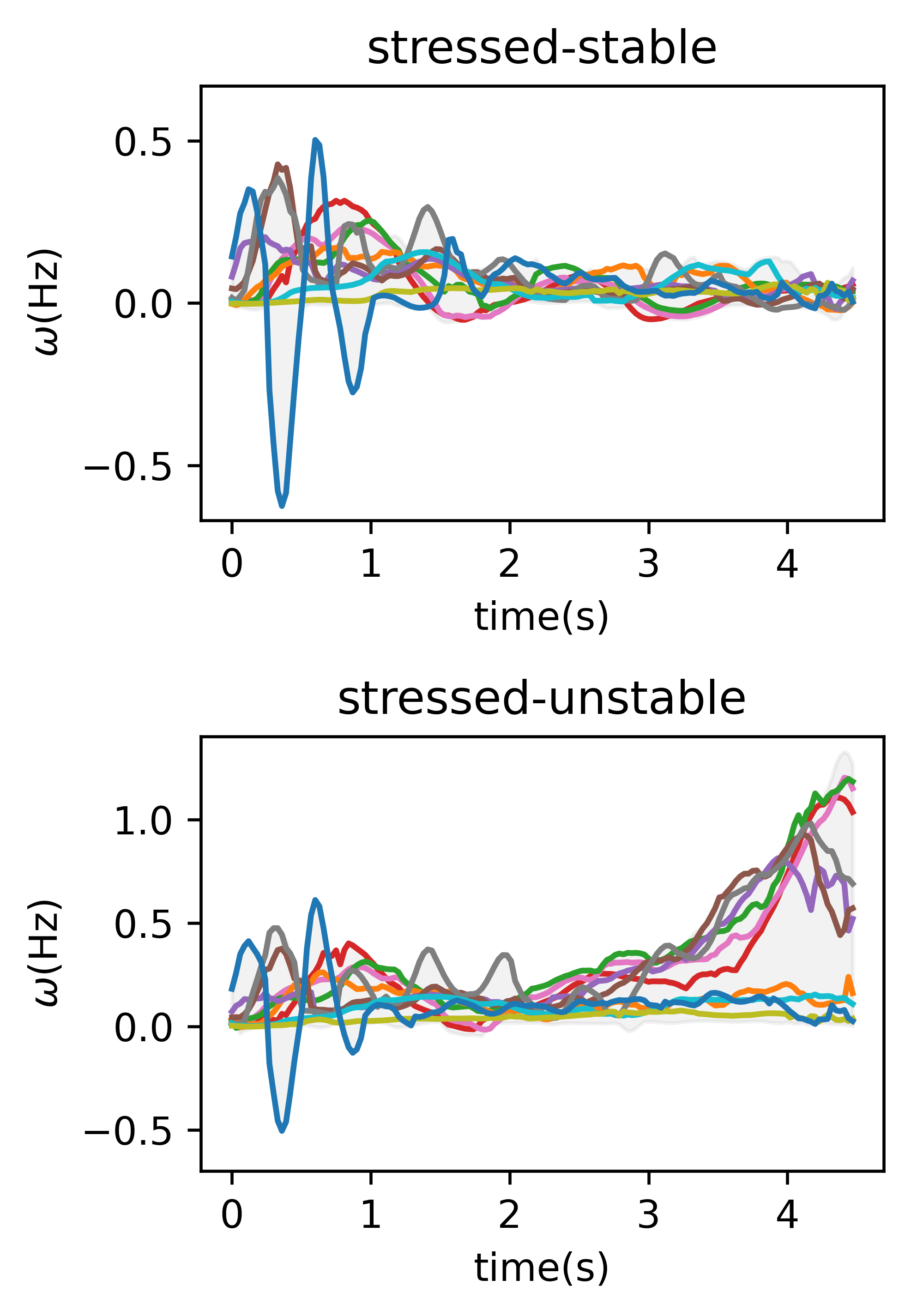}%
}
\hfil
\subfloat[ $k_{max,1}=4$ ]{\includegraphics[width=0.3\textwidth]{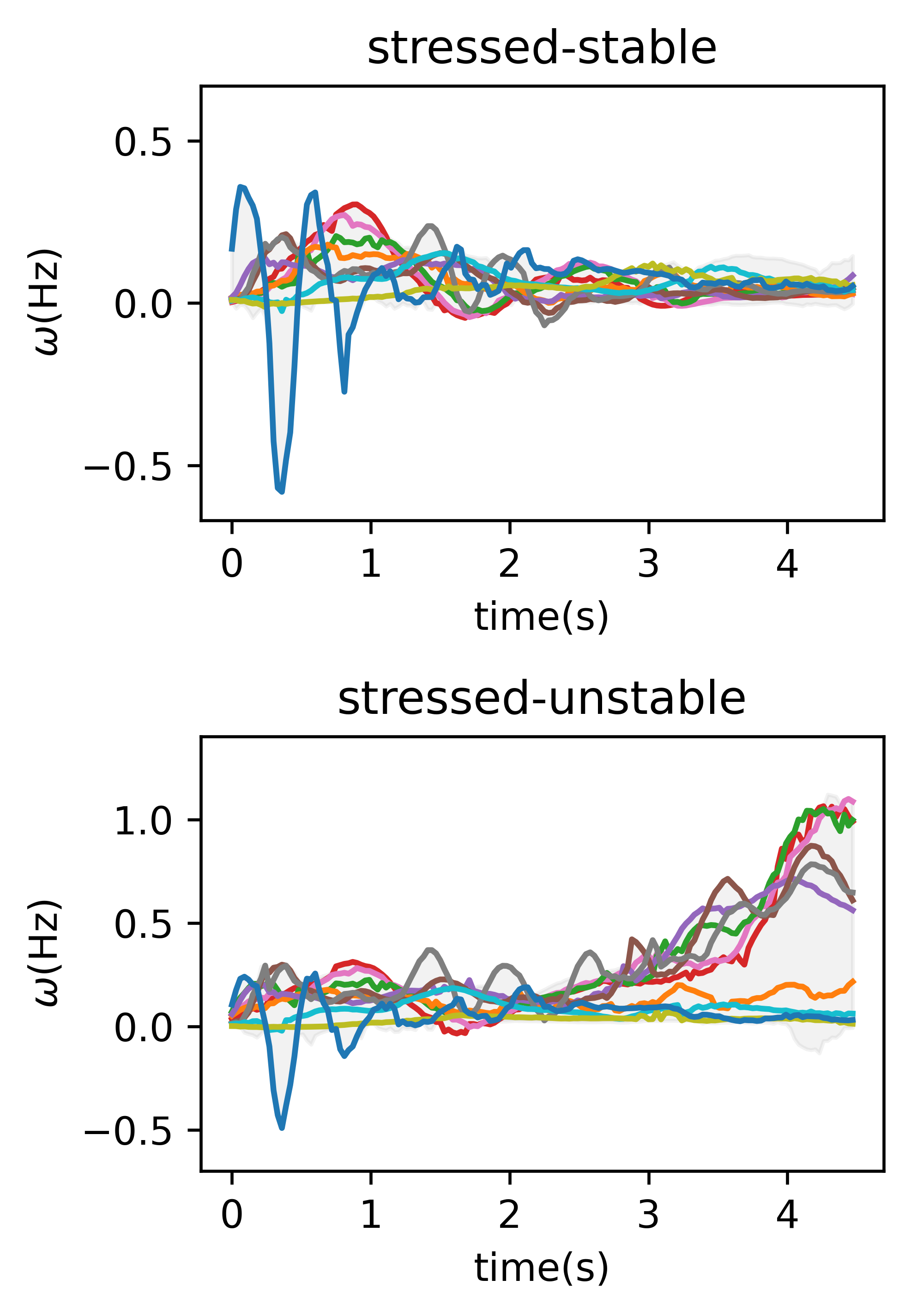}%
}
\hfil
\subfloat[ $k_{max,1}=6$ ]{\includegraphics[width=0.3\textwidth]{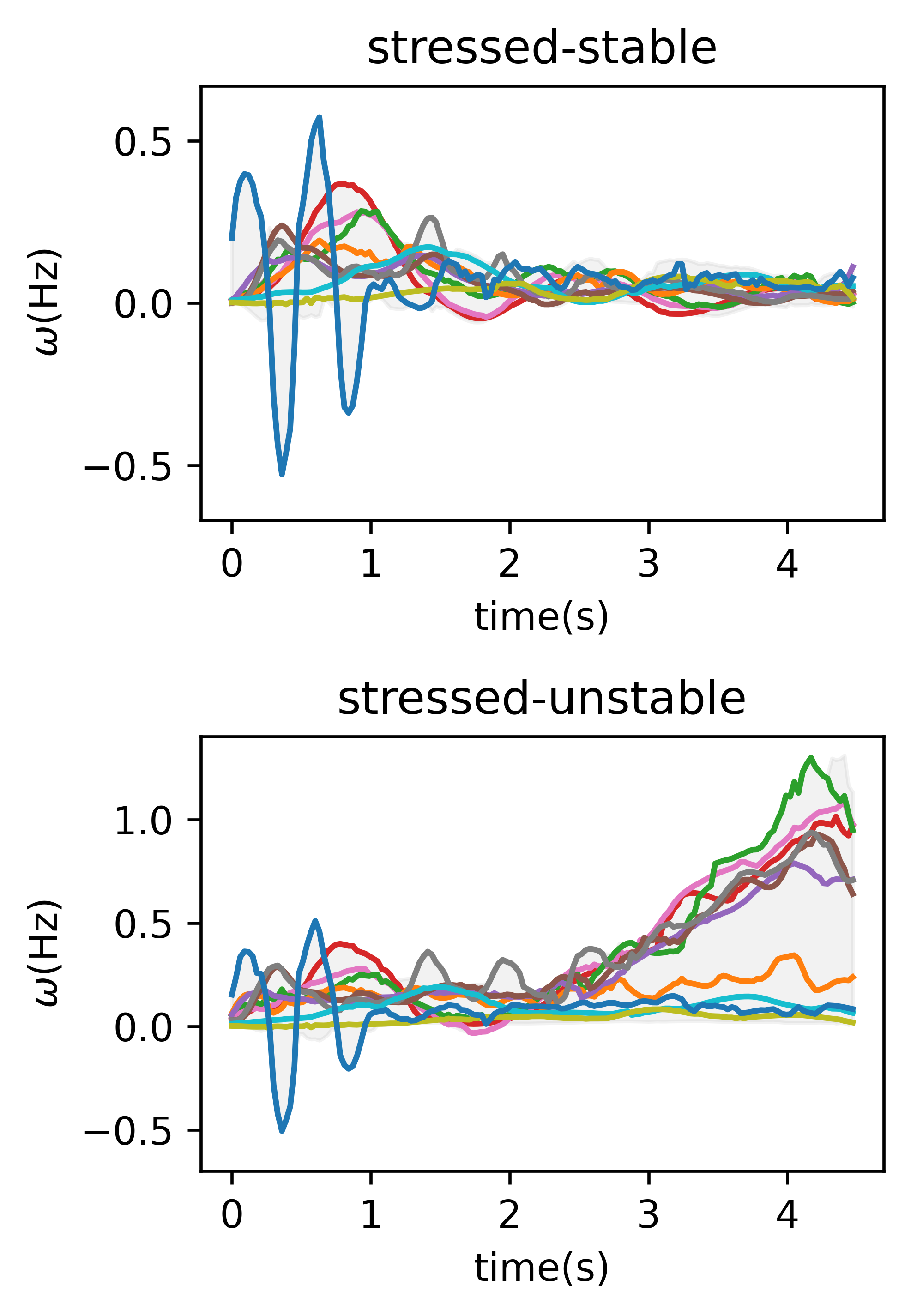}%
}
\hfil
\subfloat[ $k_{max,1}=8$ ]{\includegraphics[width=0.3\textwidth]{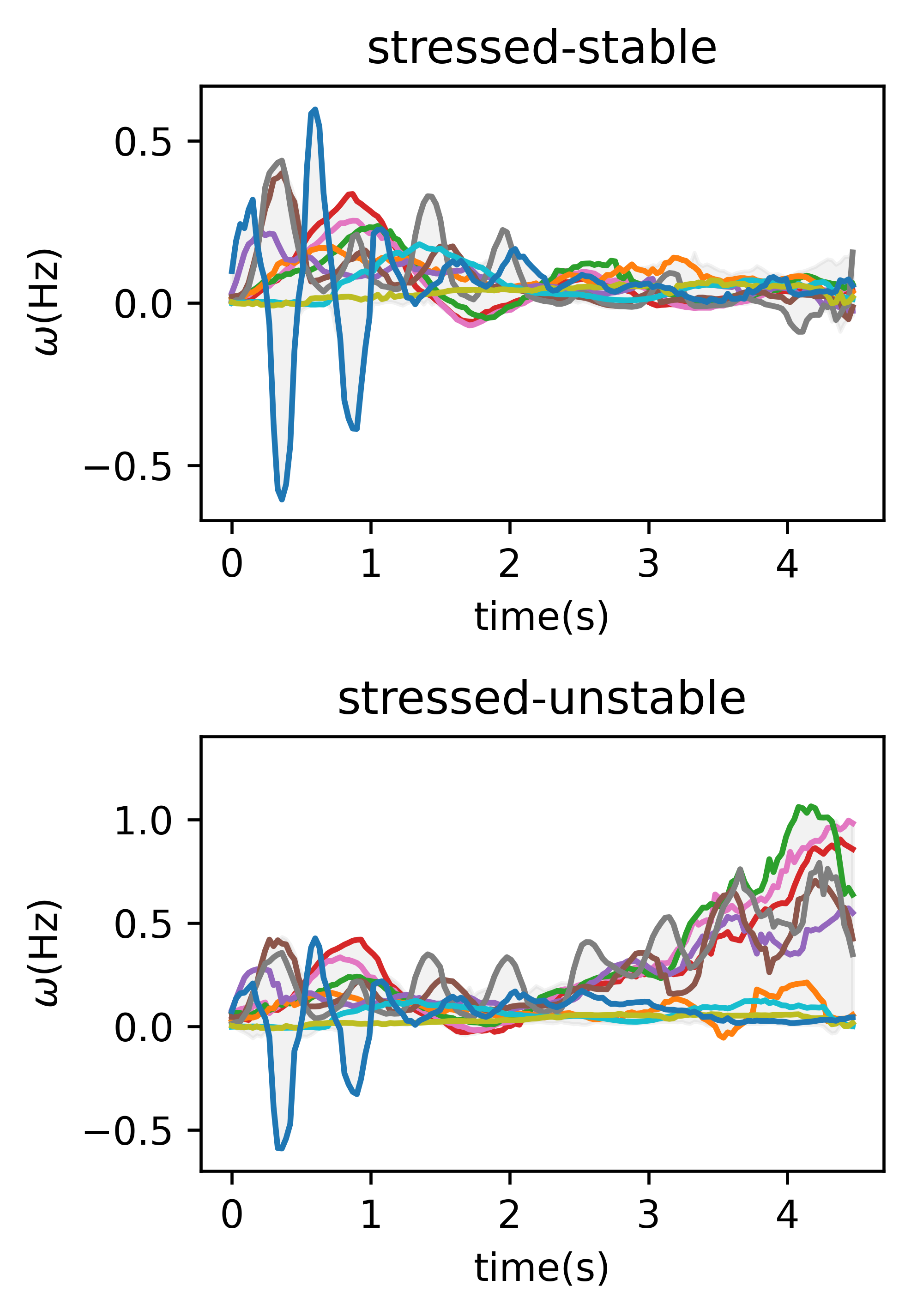}%
}
\hfil
\subfloat[ $k_{max,1}=16$ ]{\includegraphics[width=0.3\textwidth]{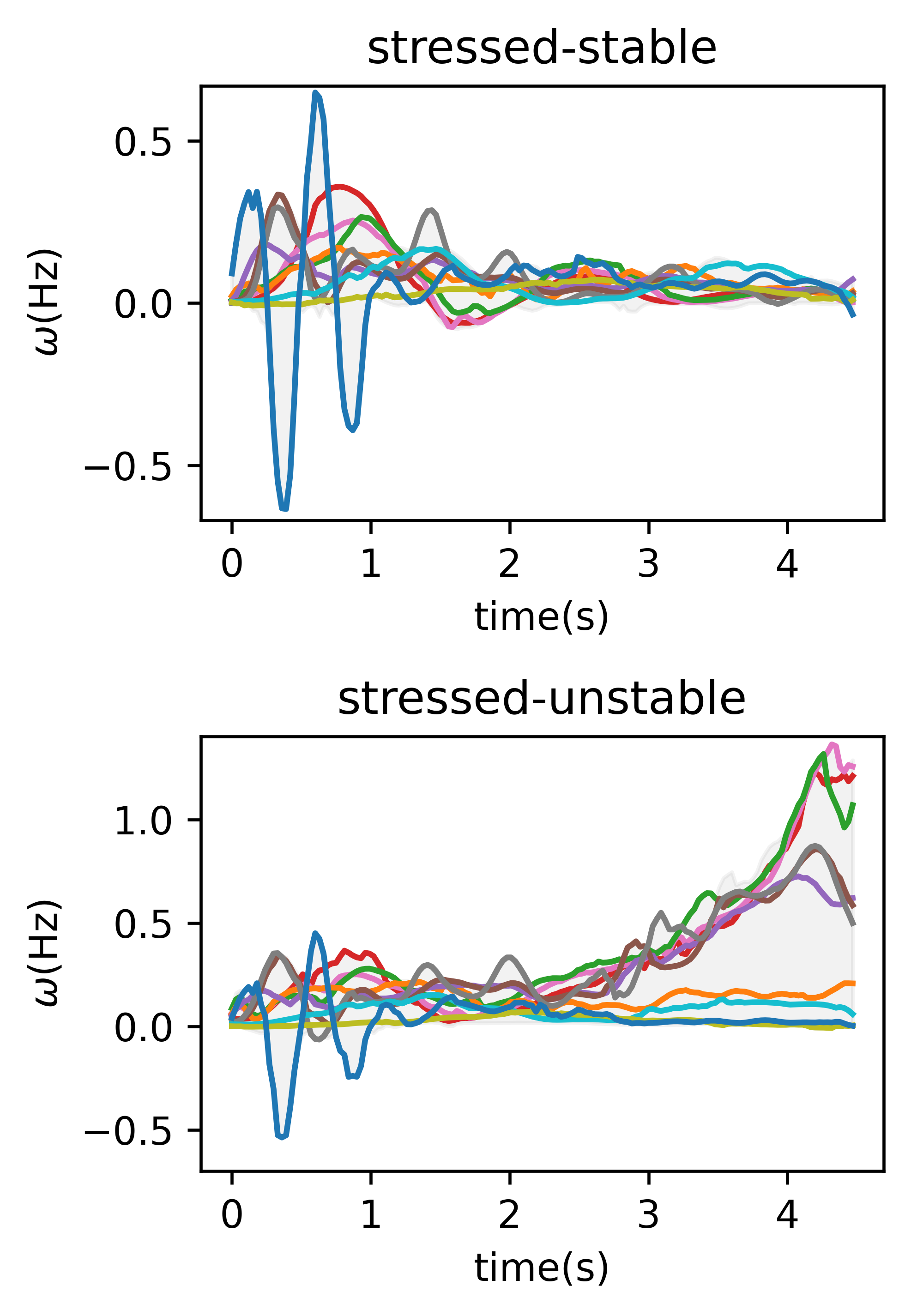}%
}
\caption{Trajectories (lines) and envelope (grey area) of the angle speed $\omega$ after a line-to-line fault between bus 75 and bus 124 and recovered at the time of 0.3s. We generate the stressed stable (the first row) and unstable case (the second row) by varying load levels. For all the three setup of $k_{\max}$, The proposed method captures the dynamics for both stable and unstable cases.}
\label{fig:dyn_stressed_r1}
\end{figure*}

\highlights{To show the effect of the low pass filter, we fix  $k_{\max,2}=3$, $k_{\max,3}=3$ and compare the performances of the trained neural networks by varying $k_{\max,1}$ to be $4$, $6$, $8$ and $16$. Note that the largest possible frequency is the Nyquist frequency where $k_{\max,1}=\tau/2$ for the trajectory of lentgh $\tau$, so the cut-off frequency is much lower than the highest frequency component (where $k_{\max,1}=75$ for $\tau=150$ in this case study). The relative mean squared error (mse) on the test set, the average prediction time and the training time is shown in Table~\ref{tab:metric_filter}. 
 Increasing $k_{\max,1}$ reduces relative mse slightly but greatly increases the prediction time and the training time. Especially, the case without the low-pass filter has the relative mse slightly lower than the case  $k_{\max,1}=6$, but the prediction and training time is 1.35 and 2.12 times as long as the value of the case $k_{\max,1}=6$, respectively.
 The prediction of the angle speed $\omega$ for stressed stable and stressed unstable cases corresponding to  the different setups of the low-pass filters are shown in Fig.~\ref{fig:dyn_stressed_r1}. For all the setup of $k_{\max}$, the shape of aperiodic and distorted waveform are all captured. 
Since the improvement brought by increasing $k_{\max,1}$ is not large after $k_{\max,1}>6$ , we select  $k_{\max,1}=6$ to report the results. }

\begin{table*}[ht]
\renewcommand{\arraystretch}{1.}
\caption{The impact of the low-pass filter}
\label{tab:metric_filter}
\centering
\small
\begin{tabular}{c||c|c|c|c|c}
\hline \hline
&w.o. the low-pass filter & $k_{\max,1}=4$ & $\bm{k_{\max,1}=6}$
& $k_{\max,1}=8$ &  $k_{\max,1}=16$ \\
\hline
\text{relative mse} &0.0062 & 0.0076 &   $\bm{00064}$
&  0.0063 &  0.0062
 \\
 \hline
\text{prediction time} & 0.0050s & 0.0034s &    $\bm{0.0037s}$
&   0.0042s &  0.0048s
 \\
  \hline
\text{training time} & 5582.16s & 4402.18s &    $\bm{4424.33s}$
&   4431.05s & 4448.27s
 \\
\hline\hline
\end{tabular}
\end{table*}

\subsection{\highlights{Influence factors including fault-on/clear actions, fault type and fault location} \label{app: factor}}

\highlights{This subsection visualizes  the predictions for the NPCC system with different influence factors. The base case is shown in Fig.~\ref{fig:dyn_stable}, where a three-phase line fault between bus 54 and bus 103 is cleared at the time of 0.3s.  We then alter fault-on/clear actions, fault type and
fault location. 
The detailed simulation results are given below.}

\begin{figure*}[ht]
\centering
\subfloat[True trajectories (lines) and envelope (grey area) after a three-phase line fault between bus 54 and bus 103 and recovered at the time of 0.15s ]{\includegraphics[width=0.8\textwidth]{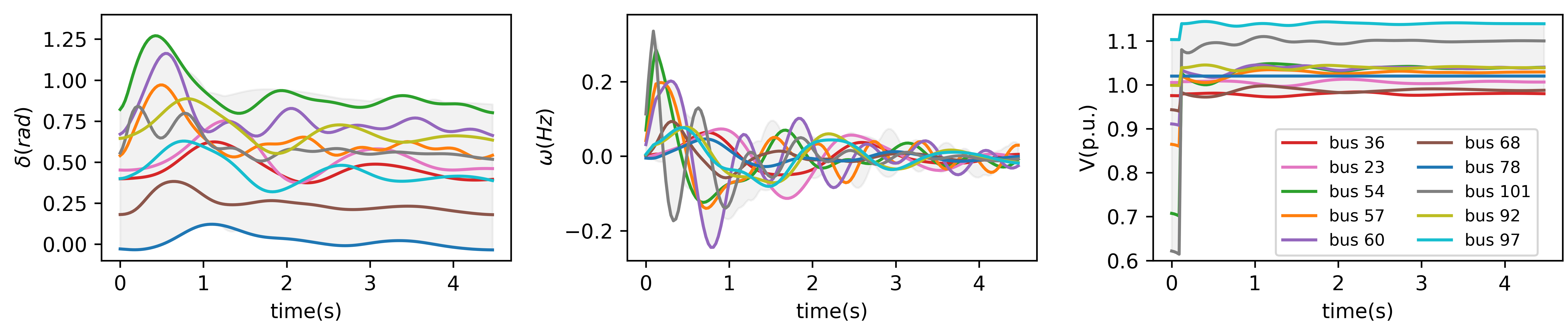}%
}
\vfil
\subfloat[Predicted trajectories (lines) and envelope (grey area) after a three-phase line fault between bus 54 and bus 103  and recovered at the time of 0.15s ]{\includegraphics[width=0.8\textwidth]{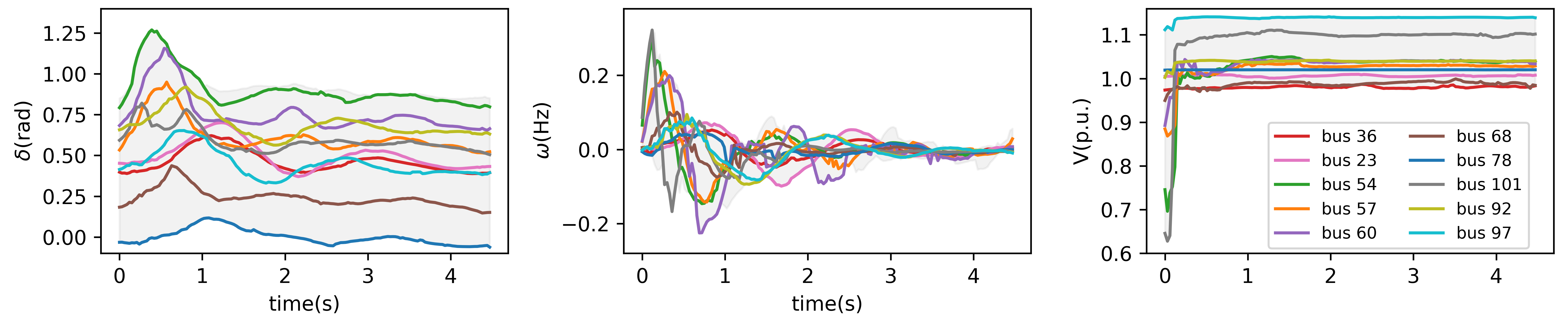}%
}
\caption{Stable dynamics of angle $\delta$ (left), frequency deviation $\omega$ (middle)  and voltage $V$ (right) in NPCC corresponding to (a)  the ground truth produced by a solver. (b) prediction of FNO. The grey area shows the envelope of the trajectories for all generator buses. Lines with different colors shows the trajectories in selected generator buses. The proposed method predict both the magnitude and oscillations accurately.}
\label{fig:dyn_stable_Line147time0.15Fault0}
\end{figure*}

\begin{figure*}[ht]
\centering
\subfloat[True trajectories (lines) and envelope (grey area) after a two-phase line fault between bus 54 and bus 103 and recovered at the time of 0.3s ]{\includegraphics[width=0.8\textwidth]{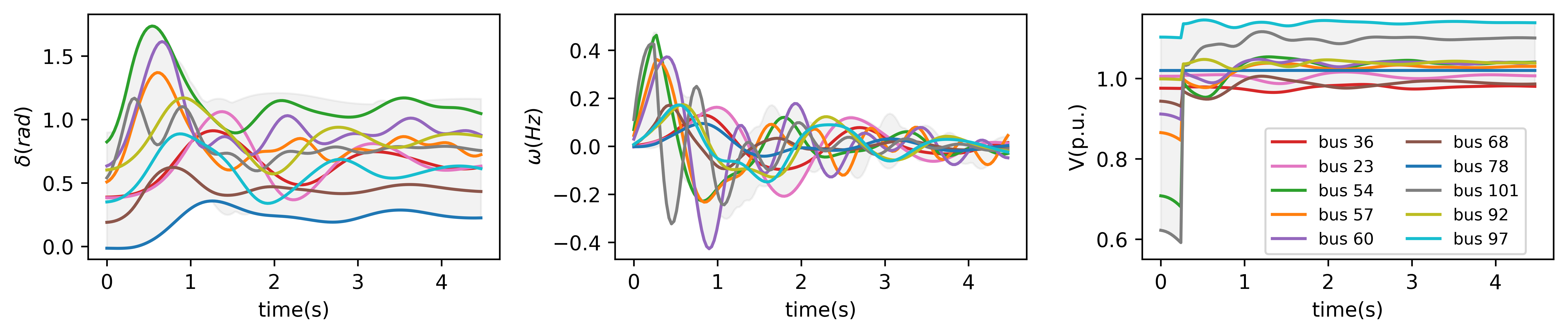}%
}
\vfil
\subfloat[Predicted trajectories (lines) and envelope (grey area) after a two-phase line fault between bus 54 and bus 103  and recovered at the time of 0.3s ]{\includegraphics[width=0.8\textwidth]{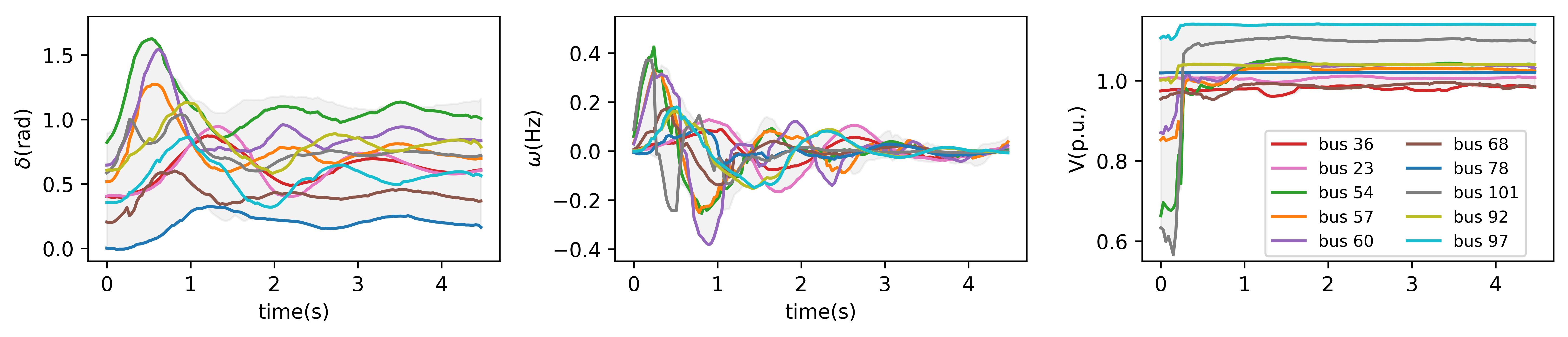}%
}
\caption{Stable dynamics of angle $\delta$ (left), frequency deviation $\omega$ (middle)  and voltage $V$ (right) in NPCC corresponding to (a)  the ground truth produced by a solver. (b) prediction of FNO. The grey area shows the envelope of the trajectories for all generator buses. Lines with different colors shows the trajectories in selected generator buses. The proposed method predict both the magnitude and oscillations accurately.}
\label{fig:dyn_stable_Line147time0.3Fault3}
\end{figure*}

\begin{figure*}[ht]
\centering
\subfloat[True trajectories (lines) and envelope (grey area) after a two-phase line fault between bus 75 and bus 124 and recovered at the time of 0.3s ]{\includegraphics[width=0.8\textwidth]{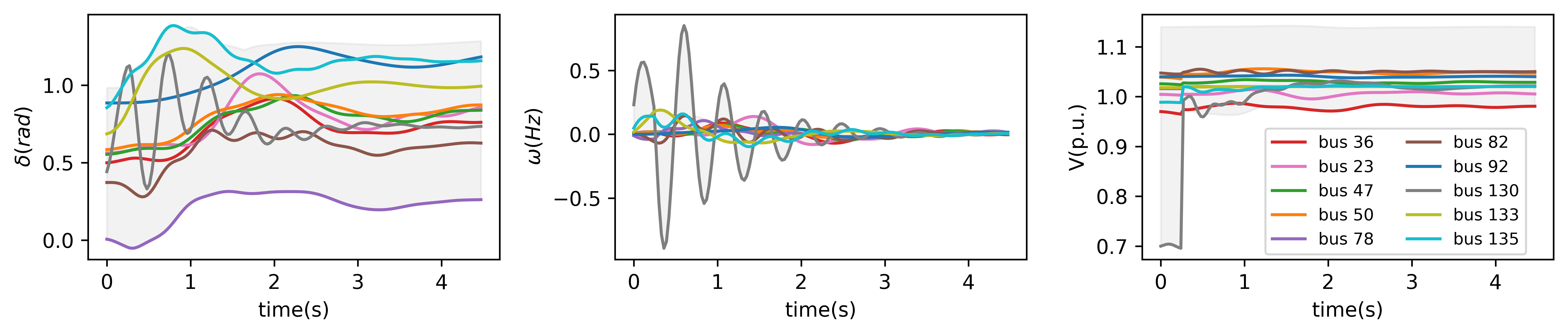}%
}
\vfil
\subfloat[Predicted trajectories (lines) and envelope (grey area) after a two-phase line fault between bus 75 and bus 124  and recovered at the time of 0.3s ]{\includegraphics[width=0.8\textwidth]{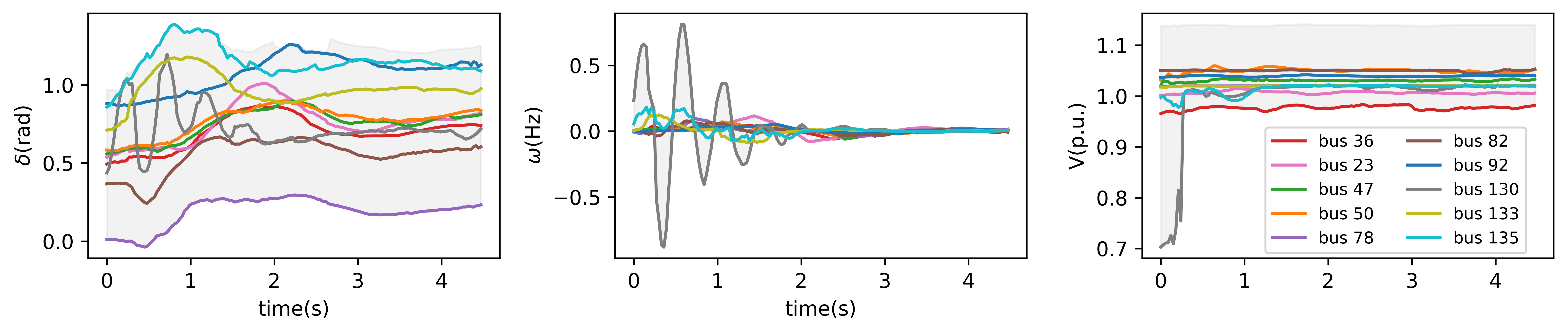}%
}
\caption{Stable dynamics of angle $\delta$ (left), frequency deviation $\omega$ (middle)  and voltage $V$ (right) in NPCC corresponding to (a)  the ground truth produced by a solver. (b) prediction of FNO. The grey area shows the envelope of the trajectories for all generator buses. Lines with different colors shows the trajectories in selected generator buses. The proposed method predict both the magnitude and oscillations accurately.}
\label{fig:dyn_stable_Line195time0.3Fault0}
\end{figure*}

\begin{enumerate}
   \item Altering fault-on/ clear actions
   
    \highlights{ In this case, we alter the fault-clear action from the time of 0.3s to 0.15s after the fault. The performance is shown in Fig.~\ref{fig:dyn_stable_Line147time0.15Fault0}, where the largest frequency deviation and the duration of voltage drop are reduced compared with Fig.~\ref{fig:dyn_stable}. Hence, the proposed method captures the difference brought by different fault-clear time.}

    \item Altering  fault type

    \highlights{  In this case, we alter the fault type from a three-phase line fault to a two-phase line fault. The performance is shown in Fig.~\ref{fig:dyn_stable_Line147time0.3Fault3}, where the major difference 
     compared with Fig.~\ref{fig:dyn_stable} is the lower magnitude of angles because of the slightly different rate-of-change of the angle speed   
     compared with Fig.~\ref{fig:dyn_stable}. Hence, the proposed method captures the difference brought by different types of faults.}

\item Altering the fault location

     \highlights{ In this case, we alter the fault location to the line  between bus
    75 and bus 124. The performance is shown in Fig.~\ref{fig:dyn_stable_Line195time0.3Fault0}, where the dynamics are very different  
     compared with Fig.~\ref{fig:dyn_stable}. The proposed method predicts the different shapes and magnitudes of oscillations. 
     Hence, the proposed method captures the difference brought by different location of faults.}
\end{enumerate}
Therefore, the proposed architecture handles all the factors well and predicts both the magnitude and oscillations accurately. 

\end{document}